\documentclass[10pt,journal,compsoc]{IEEEtran}

\usepackage{graphicx}
\usepackage[cmex10]{amsmath}
\usepackage{amsfonts}
\usepackage{amssymb}
\usepackage{mathrsfs}
\usepackage{bm}
\usepackage{algorithmic}
\usepackage{algorithm}
\usepackage{array}
\usepackage{mdwmath}
\usepackage{mdwtab}
\usepackage[caption=false,font=footnotesize]{subfig}
\usepackage{fixltx2e}
\usepackage{stfloats}
\usepackage{footnote}
\usepackage{tabularx}
\usepackage{multirow}
\usepackage{color}
\usepackage{textcomp}
\usepackage{subfig}

\newcounter{tempeqncnt}

\newtheorem{definition}{Definition}
\newtheorem{proposition}{Proposition}
\newtheorem{theorem}{Theorem}
\newtheorem{lemma}{Lemma}

\newtheorem{remark}{Remark}
\newtheorem{assumption}{Assumption}

\usepackage{color}

\title{Incentive Design in Peer Review: \\Rating and Repeated Endogenous Matching
}

\author{Yuanzhang Xiao, Florian D\"{o}rfler, and Mihaela van der Schaar
\thanks{Xiao and van der Schaar are with Department of Electrical Engineering, UCLA.
        Emails: yxiao@seas.ucla.edu  and mihaela@ee.ucla.edu }%
\thanks{D\"{o}rfler is with Automatic Control Laboratory, ETH Zurich, Switzerland. Email: dorfler@ethz.ch}
}

\begin{document}

\maketitle
\thispagestyle{empty}
\pagestyle{empty}

\begin{abstract}
Peer review (e.g., grading assignments in Massive Open Online Courses (MOOCs), academic paper review) is an effective and scalable method to evaluate the products (e.g., assignments, papers) of a large number of agents when the number of dedicated reviewing experts (e.g., teaching assistants, editors) is limited. Peer review poses two key challenges: 1) identifying the reviewers' intrinsic capabilities (i.e., \emph{adverse selection}) and 2) incentivizing the reviewers to exert high effort (i.e., \emph{moral hazard}). Some works in mechanism design address \emph{pure} adverse selection using \emph{one-shot} matching rules, and \emph{pure} moral hazard was addressed in repeated games with \emph{exogenously given} and \emph{fixed} matching rules. However, in peer review systems exhibiting both adverse selection and moral hazard, one-shot or exogenous matching rules do not link agents' current behavior with future matches and future payoffs, and as we prove, will induce myopic behavior (i.e., exerting the lowest effort) resulting in the lowest review quality.

In this paper, we propose for the first time a solution that \emph{simultaneously} solves adverse selection and moral hazard. Our solution exploits the \emph{repeated} interactions of agents, utilizes \emph{ratings} to summarize agents' past review quality, and designs matching rules that \emph{endogenously} depend on agents' ratings. Our proposed matching rules are easy to implement and require no knowledge about agents' private information (e.g., their benefit and cost functions). Yet, they are effective in guiding the system to an equilibrium where the agents are incentivized to exert high effort and receive ratings that precisely reflect their review quality. Using several illustrative examples, we quantify the significant performance gains obtained by our proposed mechanism as compared to existing one-shot or exogenous matching rules.
\end{abstract}

\section{Introduction}

\label{sec:intro}
Peer review serves as an effective and scalable method for performance evaluation in systems where the products to evaluate significantly outnumber the dedicated reviewing experts. One example of such systems is Massive Open Online Courses (MOOCs), where the number of students enrolled in a course is in the order of tens of thousands and by far exceeds the number of teaching assistants \cite{Daniel}--\cite{Wainwright}. Another example is academic paper review, where the number of papers submitted to a journal by far exceeds the number of (associate) editors. Since the proposed mechanism can be applied to general peer review settings, we keep the discussion in this paper general.\footnote{We studied explicitly the academic paper review system in the preliminary conference version \cite{XiaoDorfler_Allerton} of this paper.}

Peer review systems pose two key challenges. First, the reviewers have different \emph{intrinsic capabilities} (e.g., their review quality functions, benefit and cost functions), which are \emph{unknown}. Hence, one challenge is how to identify their \emph{unknown} intrinsic capabilities; this is known in the game theory literature as the adverse selection problem. Second, the reviewers can choose to exert different levels of (costly) \emph{effort} (e.g., time and energy spent in reviewing), which is \emph{unobservable}. Hence, the other challenge is how to incentivize reviewers to exert high effort; this is known in the game theory literature as the moral hazard problem. A reviewer's ultimate review quality is determined by her intrinsic capabilities \emph{and} effort. If the capabilities are unknown but the effort is observable (i.e. pure adverse selection), there is hope to identify their capabilities through mechanism design. If the effort is unobservable but the capabilities are known (i.e. pure adverse selection), there is hope to incentivize high effort through social norms. However, in the presence of both adverse selection and moral hazard, the problem becomes significantly more challenging. In fact, no existing work has addressed this problem systematically.

A natural candidate for solving the pure adverse selection problem is to use matching mechanisms \cite{Shapley}--\cite{RothSonmez}. Matching mechanisms aim to efficiently allocate resources (e.g., hospitals, or reviewers in our setting) to agents (e.g., medical students) or their products (e.g., assignments or papers in our setting). Existing works on matching mechanisms assume that the quality of resources depend only on the types of the agents who provide and receive the resource, but not on the providers' effort. In other words, there is no moral hazard problem. As a result, they focus on \emph{one-shot} interactions and design \emph{one-shot} matching rules (i.e., each agent is matched only once).\footnote{There are works on dynamic matching (see representative work \cite{Johari_DynamicMatching}). However, matching is called dynamic due to the dynamic arrival and departure of the agents. Each agent is still matched only once.} However, their assumption does not hold in peer review systems, where the review quality depends crucially on the reviewers' effort. We prove that under one-shot matching rules, agents will behave myopically by choosing the lowest effort (i.e., free-riding), because their current effort does not affect future matches and their future payoffs. Hence, the system performance (in terms of the total review quality) is the worst since one-shot matching does not address the moral hazard problem.

One way to address the pure moral hazard problem is to use social norms \cite{Kandori_SocialNorm}, where a central agency assigns the agents with ratings that summarize their past behavior and recommends a ``norm'' (i.e., desired behavior) that rewards agents with good ratings and punishes those with bad ratings. In this way, the agents are incentivized to conform with the social norm (e.g., exert high effort in our setting), even when they are randomly matched to each other based on some \emph{exogenous} matching rule. However, existing works on social norms assume that the agents are \emph{homogenous}. This assumption does not hold in peer review systems, because different reviewers have different intrinsic capabilities. Ideally, the central agency should recommend different norms to agents of different capabilities; in practice, it cannot do this since the capabilities are unknown. In summary, existing works in social norms \cite{Kandori_SocialNorm}--\cite{XiaoMihaela_TEAC} do not deal with the adverse selection problem that is present in peer review.

This paper proposes the \emph{first} mechanism to \emph{simultaneously} solve the adverse selection and moral hazard problems in peer review. Our proposed mechanism exploits the repeated interaction among agents (i.e., by submitting multiple assignments or papers over time), and assigns the agents with ratings, which are summaries of their past review quality. Unlike the works on social norms \cite{Kandori_SocialNorm}--\cite{XiaoMihaela_TEAC}, we do not recommend desired behavior (i.e., a social norm) to the agents, because they have unknown, different capabilities and thus computing a recommended social norm is impossible. Instead, we propose rules for \emph{repeated} matching that \emph{endogenously} depend on agents' ratings. Unlike existing one-shot \cite{Shapley}--\cite{RothSonmez} or exogenous \cite{Kandori_SocialNorm}--\cite{XiaoMihaela_TEAC} matching rules, our proposed repeated endogenous matching rules provide strong incentives for agents to exert high effort, because the agents' behaviors affect their future ratings and hence, their future matches and future payoffs. 

We provide design guidelines for endogenous matching rules that are easy to implement without knowledge of agents' private information (e.g., their benefit, review quality, and cost functions), yet powerful enough to guide the system to desirable equilibria. In particular, in the equilibrium the agents find it in their self-interest to exert high effort, and receive ratings that truly reflect their capabilities. We also provide case studies on specific matching rules with different reward/punishment schemes. We show that different reward/punishment schemes lead to different optimal matching rules, which stresses the importance of tailoring matching rules to reward/punishment schemes. Simulation results demonstrate large performance improvement over existing matching rules.

In the following, we discuss related works in Section~\ref{sec:related}. Then we describe the model and formulate the design problem in Section~\ref{sec:model}. We study general matching rules in Section~\ref{sec:convergence}, and the baseline matching rule and its extensions in Section~\ref{sec:design}. Section~\ref{sec:simulation} demonstrates the efficiency of our proposed mechanisms. Finally, Section~\ref{sec:conclusion} concludes the paper.

\section{Related Works}\label{sec:related}

\subsection{Pure Adverse Selection}

The pure adverse selection problem is the focus of a huge literature on matching in resource allocation (e.g., allocation of schools to applicants \cite{Shapley}\cite{AbdulkadirogluSonmez}) and exchange (e.g., kidney exchange \cite{RothSonmez}). These works ignore the moral hazard problem. In particular, they do not consider ``effort''. Once an agent (e.g., an applicant) is matched to another (e.g., a school), the benefit (obtained by this applicant) is fixed. In contrast, in our work, the review quality depends crucially on the reviewer's effort. This additional moral hazard problem, when ignored, will significantly degrade the system performance (in terms of the total review quality).

Since these works \cite{Shapley}--\cite{RothSonmez} ignore moral hazard, their matching rules are one-shot (i.e., match each agent only once). In contrast, our matching rules are repeated and changing over time based on agents' ratings. Hence, our matching rules can incentivize agents to exert high effort to obtain better ratings and thus, favorable future matches.

\subsection{Pure Moral Hazard}
The pure moral hazard problem has been studied in repeated game theory, where anonymous agents are randomly matched to interact with each other \cite{Kandori_SocialNorm}--\cite{XiaoMihaela_TEAC}, as in our work. However, these works \cite{Kandori_SocialNorm}--\cite{XiaoMihaela_TEAC} focus on the \emph{pure} moral hazard problem, and ignore the adverse selection problem by assuming \emph{homogeneous} agents. In this work, we assume \emph{heterogeneous} agents and deal with both the moral hazard and adverse selection problems. In \cite{Kandori_SocialNorm}--\cite{XiaoMihaela_TEAC}, due to the homogeneity of agents, binary ratings are usually sufficient to identify whether a player has behaved well or badly. In contrast, in this work the rating is continuous, such that the rating mechanism can identify not only whether a player has behaved well or badly, but also its review quality.

Another key difference is that we \emph{design} matching rules that \emph{endogenously} depend on agents' ratings and directly affect their incentives, while the matching rules in \cite{Kandori_SocialNorm}--\cite{XiaoMihaela_TEAC} are \emph{fixed} and \emph{exogenously given}. In our setting, we will prove that the latter type of matching rules will result in the lowest review quality in the equilibrium.

\subsection{Other Works on Peer Review}
There are other works on peer review systems but with different problems to solve. For example, there are works focusing on how to aggregate reviewers' scores/ratings to obtain a final score/rating that accurately reflects the true quality of the assignments  (in MOOCs \cite{NgKoller}--\cite{Wainwright}), the proposals (in NSF proposal reviewing \cite{Procaccia}), or the papers (in academic peer review \cite{Bianchi_PeerReview}). In contrast, our focus is to incentivize reviewers to exert high effort levels.

\section{Model}\label{sec:model}

\subsection{Basic Setup}
Consider a peer review system with a set $\mathcal{N}=\{1,\ldots,N\}$ of $N$ agents. Each agent has its products reviewed by the other agents. An agent benefits from the review by its reviewer, and exerts effort in reviewing others' products. A designer (e.g., the instructor in MOOCs)  aims to design a mechanism that incentivizes the reviewers to produce high-quality reviews. The mechanism includes two parts: {\em (i)} the rating mechanism that assigns and updates a rating $\theta_i \in \mathbb{R}_+$ for each agent $i$, and {\em (ii)}  the matching rule that matches agents with reviewers (possibly based on the ratings). In the following, we write the rating profile, namely the ratings of every agent, as $\bm{\theta}=(\theta_1,\ldots,\theta_N)$. The rating profile is known only to the designer. We define the \textbf{rating distribution}, denoted by a vector $d(\bm{\theta})$, as the ordered (from high to low) list of all the ratings. The rating distribution $d(\bm{\theta})$ does not count multiple agents with the same rating. For example, if the rating profile is $\bm{\theta}=(3,5,5,3)$, the rating distribution will be $d(\bm{\theta})=(5,3)$. Write $K$ as the number of distinct ratings in $\bm{\theta}$ (i.e., the dimension of the vector $d(\bm{\theta})$), and $k_i$ as agent $i$'s ranking (i.e., ordered position) in the rating distribution. In the above example, we have $K=2$, $k_1=k_4=2,~k_2=k_3=1$. Although $K$ and $k_i$ depend on $\bm{\theta}$, we write them simply as $K$ and $k_i$ for notational simplicity without causing confusion. Denote the $k_i$th element of the rating distribution by $d(\bm{\theta})_{k_i}$. Then we have $\theta_i=d(\bm{\theta})_{k_i}$. Finally, notice that the rating distribution does not disclose any information about the identities of the agents.

Note, importantly, that an agent's rating indicates its review quality, not the quality of its product.

Time is slotted into $t=0,1,2,\ldots$. In each time slot $t$, the entities in the system moves in the following order:\footnote{Throughout the paper, the superscript $(\cdot)^\prime$ on a function refers to the derivative, and the superscript $(\cdot)^{t}$ refers to the variable under consideration at time point $t \in \mathbb Z_{+}$.}
\begin{itemize}
\item The designer publishes the rating distribution $d(\bm{\theta})$, and informs agent $i$ of its rating $\theta_i$ and its ranking $k_i$.
\item Each agent submits its product to review.
\item The designer matches each agent $i$'s product to other agent(s) for review based on a probabilistic \textbf{matching rule} $m_{k_i k_j}: (d(\bm{\theta})_{k_i},d(\bm{\theta})_{k_j}) \mapsto [0,1]$. 
The matching rule determines the probability $m_{k_i k_j}(d(\bm{\theta})_{k_i},d(\bm{\theta})_{k_j})$ that the agent with the $k_i$th highest rating is matched to the reviewer with the $k_j$th highest rating. From the definition we can see that the matching does not depend on agents' identities.
\item Each reviewer $j$ exerts an \textbf{effort level} $e_j^t \in [0, e_j^{\rm max}]$, where $e_j^{\rm max}$ is $j$'s maximum effort level. Reviewer $j$'s \textbf{review quality} then depends on its effort as $q_j(e_j^t)$, where $q_j: \mathbb{R}_+ \rightarrow \mathbb{R}_+$ is the review quality function.
\item Each agent $i$ receives \textbf{benefit} $b_i(q_j(e_j^t))$ from reviewer $j$'s review, where $b_i: \mathbb{R}_+ \rightarrow \mathbb{R}_+$ is $i$'s benefit function, and incurs a \textbf{cost} of $c_i(e_i^t)$ for reviewing a product, where $c_i: \mathbb{R}_+ \rightarrow \mathbb{R}_+$ is $i$'s cost function.
\item Each agent $i$ sends a \textbf{report} $r_i^t$ about the reviewer to the designer. We assume that the report accurately reflects the reviewer's review quality, namely $r_i^t = q_j(e_j^t)$. For examples, in MOOCs the report can be made accurate by comparing the grading with the true answers to the assignments (posted after the submission of the assignments). 
\item The designer updates the agents' ratings according to the \textbf{rating update rule} $\pi: (\theta_j^t, r_i^t) \mapsto \theta_j^{t+1}$. For fairness, the rating update rule is identical for all reviewers, and is given by a convex combination of the reviewer's old rating and the report about its review quality with a constant step size $\mu \in (0,1)$:\footnote{Note that under the assumption that $j$'s review quality $q_j(e_j)$ is perfectly observed, it does not matter how many reports about $j$'s review quality are received.}
\begin{eqnarray}\label{eqn:RatingUpdateRule}
\!\!\!\!\!\!\!\!\pi(\theta_j^t, r_i^t)= \left\{ \begin{array}{cl} (1-\mu) \cdot \theta_j^t + \mu \cdot r_i^t, & j~\text{has~reviewed} \\ \theta_j^t, & \text{otherwise} \end{array} \right. \!\!\!\!\!\!\!\!
\end{eqnarray}
\end{itemize}

We make the following remarks on the agents' ratings. Each agent $i$ has a maximum effort level $e_i^{\rm max}$, and hence has a maximum review quality $q_i(e_i^{\rm max})$. Since the new rating is the convex combination of the old rating and the review quality, given any initial rating $\theta_i^0$, agent $i$'s rating can only be in the interval $\left[0, \theta_i^{\rm max}\right]$, where $\theta_i^{\rm max} \triangleq \max\left\{\theta_i^0, q_i(e_i^{\rm max})\right\}$. In other words, the possible ratings of each agent $i$ are contained in the compact set $\left[0, \theta_i^{\max}\right]$.

Throughout the paper, we make the following reasonable and standard assumptions on the monotonicity, convexity, concavity, and differentiability of our functions.

\begin{assumption}[Cost, Review Quality, and Benefit]\label{assumption:Cost-ReviewQuality}
Each agent $i$'s cost function $c_i(\cdot)$, review quality function $q_i(\cdot)$, and benefit function $b_i(\cdot)$ satisfy the following:
\begin{itemize}
\item The cost $c_i(\cdot)$ is strictly convex, strictly increasing, and twice continuously differentiable in effort $e_i$. In addition, $c_i^\prime(0)=0$.
\item The review quality $q_i(\cdot)$ is concave, strictly increasing, and twice continuously differentiable in effort $e_i$. In addition, $q_i^\prime(0)$ exists and is bounded.
\item The benefit $b_i(\cdot)$ is strictly increasing, concave, and continuously differentiable in review quality $q_j$. In addition, $b_i^\prime(0)$ exists and is bounded.
\item We normalize $c_i(0)=0$, $q_i(0)=0$, and $b_i(0)=0$.
\end{itemize}
\end{assumption}

\subsection{Information -- Who Knows What}

\subsubsection{The designer}

The designer receives reports $r_i$ of review quality, and keeps the rating $\theta_i^t$ for each agent $i$ at each time slot $t$. Hence, the designer knows the identity of the agent at the $k_i$th position of the rating distribution. However, it does not know the review quality functions $q_i(\cdot)$, the benefit functions $b_i(\cdot)$, or the cost functions $c_i(\cdot)$. 

\subsubsection{Each agent $i$}
Each agent $i$ knows its own review quality function $q_i(\cdot)$, benefit function $b_i(\cdot)$, and cost function $c_i(\cdot)$, but does not know the above functions of the other agents. It knows the matching rule $m$ and the rating update rule $\pi$. It also knows its own rating $\theta_i^t$, the rating distribution $d(\bm{\theta})^{t}$, and its position in the rating distribution $k_i^t$, but does not know the others' ratings or the identity of its reviewer.

\subsection{Payoffs and Equilibrium}

In each time slot $t$, agent $i$'s expected payoff is its expected benefit from the reviewing of its product minus the expected cost of reviewing other agents' products. We write agent $i$'s expected payoff as $u_i(m, \theta_i, d(\bm{\theta}), \bm{e})$, which depends on the matching rule $m$, its own rating $\theta_i$, the rating distribution $d(\bm{\theta})$, and all the agents' effort levels $\bm{e} \triangleq (e_1, \ldots, e_N)$. The \emph{expected payoff} can be calculated as the expected benefit minus the expected cost:
\begin{eqnarray}
u_i\left(m, \theta_i, d(\bm{\theta}), \bm{e}\right)\!\!\!\!\!
&=& \!\!\!\!\!\sum_{k_j \neq k_i} m_{k_i k_j}\left( d(\bm{\theta})_{k_i}, d(\bm{\theta})_{k_j} \right) \cdot b_i\left(q_j\left(e_j\right)\right) \nonumber \\
&& \!\!\!\!\!\!\!\!\!\!\!\!\!\! - \left[\sum_{k_j \neq k_i} { m_{k_j k_i}\left(d(\bm{\theta})_{k_j},d(\bm{\theta})_{k_i}\right)}\right] \cdot c_i(e_i). 
\end{eqnarray}

Each agent $i$ aims to choose a sequence of effort levels over time to maximize the \emph{discounted average of expected payoffs}, i.e., to solve the {\em dynamic} optimization problem below:
\begin{eqnarray}\label{eqn:original_optimization}
\!\!\max_{ \left\{ e_i^{t} \in \left[0, e_i^{\rm max}\right] \right\}_{t=0}^\infty} 
\!\!\!\!\mathbb{E} \left\{ (1-\delta_i) \sum_{t=0}^\infty \delta_i^{t} 
u_i\left(m, \theta_i^t, d(\bm{\theta}^t), e_i^t, \bm{e}_{-i}^t\right) \right\},\!\!\!\!
\end{eqnarray}
where $\bm{e}_{-i}^t$ is the effort levels chosen by all the agents other than $i$ at time $t$, and $\delta_i \in [0,1)$ is agent $i$'s \emph{discount factor}. An agent's discount factor reflects its patience. We take the expectation $\mathbb{E}\{\cdot\}$ because the rating update is random, namely an agent's rating is either updated or kept the same depending on whether it has reviewed a product.

Note that the optimization problem \eqref{eqn:original_optimization} is very hard, if not impossible, to solve. The difficulty lies in the couplings of one agent's decisions over time and with other agents' decisions. First, the agent's current decision (i.e., effort level) affects not only its current payoff (through the cost), but also its future ratings and hence future payoffs. Second, the agent's payoff is affected by the others' decisions (through the benefit). 
However, since an agent has no knowledge about the others, it cannot predict the others' decisions and the evolution of rating distributions. In summary, an agent cannot solve the optimization problem \eqref{eqn:original_optimization} due to computational complexity and lack of knowledge. 

We propose a realistic behavioral model for the agents. To choose the optimal effort level at each time $t$, each agent $i$ holds a {\em conjecture} that its future value $\mathbb{E} \left\{ (1-\delta_i) \sum_{\tau=t}^\infty \delta_i^{\tau-t} 
u_i(m, \theta_i^\tau, d^\tau, e_i^\tau, \bm{e}_{-i}^\tau) \right\}$ (i.e., its discounted average payoff after time $t$) is the following:
\begin{eqnarray}\label{eqn:conjecture}
f_i\left( \alpha_i,\beta_i^t, \theta_i^t, d(\bm{\theta}^t), e_i^t \right) \triangleq \alpha_i  \cdot \bar{b}_i\left(\theta_i^t, d(\bm{\theta}^t), e_i^t\right) + \beta_i^t,
\end{eqnarray}
where $\bar{b}_i\left(\theta_i^t, d(\bm{\theta}^t), e_i^t\right)$ is the conjectured expected benefit of agent $i$ in time $t+1$, assuming that the others' ratings remain the same. We can calculate $\bar{b}_i\left(\theta_i^t, d(\bm{\theta}^t), e_i^t\right)$ as
\begin{eqnarray}
&&\bar{b}_i\left(\theta_i^t, d(\bm{\theta}^t), e_i^t\right) \nonumber \\
&=& \textstyle\sum_{k_j \neq k_i^{+}} \Big[ 
m_{k_i^{+} k_j}\big(\underbrace{\pi\left(\theta_i^t ,q_i(e_i^t)\right)}_{\triangleq \theta_i^{t+1}}, d(\theta_i^{t+1},\bm{\theta}_{-i}^t)_{k_j}\big) \Big. \nonumber \\
&& ~~~~~~~~~~~~~~\textstyle\Big. \cdot b_{i}\left(d(\theta_i^{t+1},\bm{\theta}_{-i}^t)_{k_j}\right) \Big], 
\end{eqnarray}
where $d(\theta_i^{t+1},\bm{\theta}_{-i}^t)$ is the new rating distribution when $i$'s rating is updated to $\theta_i^{t+1}\triangleq \pi\left(\theta_i^t ,q_i(e_i^t)\right)$ and the others' ratings remain the same, and $k_i^+$ is $i$'s new ranking of its new rating $\theta_i^{t+1}$ in the new rating distribution $d(\theta_i^{t+1},\bm{\theta}_{-i}^t)$. Note that $i$ can compute $d(\theta_i^{t+1},\bm{\theta}_{-i}^t)$ based on $\theta_i^{t+1}$ and $d(\bm{\theta}^t)$, without knowing $\bm{\theta}_{-i}^t$.

Each agent $i$ holds the conjecture \eqref{eqn:conjecture} for two reasons. First, it cannot predict the others' effort levels or future ratings. Hence, it holds a conjecture that \emph{the others' ratings remain the same}, and that the others' ratings precisely reflect their review quality, namely $d(\bm{\theta}^t)_{k_j}=q_j(e_j^t)$. Second, it conjectures that its future value is an \emph{affine function of its expected benefit}. For consistency, both of the above conjectures are required to be true in the equilibrium to be defined later.

The coefficient $\alpha_i$ reflects how ``optimistic'' an agent is about the rating mechanism. An agent with a larger $\alpha_i$ ``believes in'' the rating mechanism more, because it anticipates a higher future value given the expected benefit. The coefficient $\beta_i^t$ is updated in each time slot by agent $i$, such that the conjectured future value converges to the true future value in the equilibrium.

Then at each time $t$, each agent $i$ simply solves the following static problem for its optimal effort level $e_i^{t}$:
\begin{eqnarray}\label{eqn:BestResponse}
e_i^{t} = \arg \max_{e_i\in \left[0, e_i^{\rm max} \right]} & & \!\!\!\!\!\!\!\! (1-\delta_i) \cdot u_i\left(m, \theta_i^t, d(\bm{\theta}^t), e_i, \bm{e}_{-i}\right) \nonumber \\
&+& \delta_i \cdot f_i\left(\alpha_i,\beta_i^t, \theta_i^t, d(\bm{\theta}^t), e_i\right).
\end{eqnarray}
Note that the others' current effort levels $\bm{e}_{-i}$ only affect the benefit term in the current payoff $u_i\left(m, \theta_i^t, d(\bm{\theta}^t), e_i, \bm{e}_{-i}\right)$, which does not depend on $i$'s effort $e_i$ and can be considered as a constant. Hence, each agent $i$ has all the information needed to solve the above static optimization problem.

\begin{definition}[Conjectural Equilibrium \cite{Hahn}]
Given any matching rule $m$ and any rating update rule $\pi$, a conjectural equilibrium (CE) is a triple $\{\theta_i^*,e_i^*,\beta_i^*\}_{i \in \mathcal{N}}$ that satisfies:
\begin{itemize}
\item Incentive compatibility constraints: for all $i \in \mathcal{N}$,
\begin{eqnarray}
e_i^* = \arg\max_{e_i\in [0, e_i^{\rm max}]} & &  \!\!\!\!\!\!\!\!\!\!\! (1-\delta_i) \cdot u_i\left(m, \theta_i^*, d(\bm{\theta}^*), e_i, \bm{e}_{-i}^*\right) \nonumber \\
& & \!\!\!\!\!\!\!\! \!\!\!\!\!\!\!\! + \delta_i \cdot f_i\left(\alpha_i,\beta_i^*, \theta_i^*, d(\bm{\theta}^*), e_i\right),
\end{eqnarray}
\item Stable and correct ratings: for all $i \in \mathcal{N}$, $\theta_i^*=q_i(e_i^*)$,
\item Consistent conjectures: for all $i \in \mathcal{N}$,
\begin{eqnarray}
f_i\left(\alpha_i,\beta_i^*, \theta_i^*, d(\bm{\theta}^*), e_i^*\right) = u_i\left(m, \theta_i^*, d(\bm{\theta}^*), \bm{e}^*\right).
\end{eqnarray}
\end{itemize}
\end{definition}

In the above definition, the incentive compatibility constraints ensure that the effort level $e_i^*$ is the best response of each agent $i$. In other words, it will be in agent $i$'s self-interest to choose $e_i^*$. A CE also requires that each agent's rating truly reflects its review quality at the equilibrium effort level $e_i^*$, and hence each agent's rating is stable, namely $\pi(\theta_i^*, q_i(e_i^*)) = \theta_i^*$. Finally, a CE requires that each agent's conjecture about its future value is correct.

There may be many CEs. As a designer, it is desirable that the system will converge to a CE from any initial rating profile. The convergence is important, because the designer can distinguish the true review quality of the reviewers at the equilibrium. The choice of the matching rule plays an important role in ensuring the convergence to a CE. Aside from convergence, certain CEs are more desirable than others, as discussed in the next paragraphs.

\subsection{The Design Problem Formulation}

The designer's problem is to maximize the equilibrium review quality. We write the designer's objective as a function of the equilibrium review quality $W\left(q_1(e_1^*),\ldots,q_N(e_N^*)\right)$. Then the designer problem can be defined as
\begin{eqnarray}
&\max_{m,\pi}& W\left(q_1(e_1^*),\ldots,q_N(e_N^*)\right) \\
& s.t.        & \{\theta_i^*,e_i^*,\beta_i^*\}_{i \in \mathcal{N}}~\mathrm{ is~a~CE~under}~m, \pi. \nonumber
\end{eqnarray}

Note that the designer does {\em not} maximize the social welfare (i.e., the total benefit minus cost of the agents), because it is more natural from the designer's perspective to maximize the total review quality. The designer of the peer review system may not care about the cost of reviewing; in fact, it would like to elicit more effort from the reviewers, resulting in higher-quality reviews but higher costs.

\section{Convergence to Conjectural Equilibria}\label{sec:convergence}

In this section, we consider general matching rules, and provide important guidelines for designing the matching rules. As discussed before, we would like to have a matching rule under which the system will converge to a CE from any initial rating profile under the best response dynamics. In this way, the designer can distinguish the true quality of the reviewers in the equilibrium. Before discussing the properties of the matching rules that ensure the convergence, we first describe the best response dynamics.

At each time slot $t$, the best response dynamics consist of the following three updates:
\begin{eqnarray}\label{eqn:UpdateEffort}
e_i^t &= \displaystyle \arg\max_{e_i\in [0, e_i^{\rm max}]}  & (1-\delta_i) \cdot u_i\left(m, \theta_i^t, d(\bm{\theta}^t), e_i, \bm{e}_{-i}^{t}\right)  \nonumber \\
& & + \delta \cdot f_i\left(\alpha_i,\beta_i^t, \theta_i^t, d(\bm{\theta}^t), e_i\right);
\end{eqnarray}

\begin{eqnarray}\label{eqn:UpdateRating}
\theta_i^{t+1} = \left\{ \begin{array}{cl} (1 - \mu) \cdot \theta_i^t + \mu \cdot q_i\left(e_i^t\right) & \mathrm{if}~i~\mathrm{reviewed} \\ \theta_i^t & \mathrm{otherwise} \end{array} \right.;
\end{eqnarray}

\begin{eqnarray}\label{eqn:UpdateBelief}
\beta_i^{t+1}  = u_i\left(m,\theta_i^{t}, d(\bm{\theta}^{t}), \bm{e}^{t}\right) - \alpha_i \cdot \bar{b}_i\left(\theta_i^t, d(\bm{\theta}^t), e_i^t\right).  
\end{eqnarray}

The update of effort levels in \eqref{eqn:UpdateEffort} and the update of ratings in \eqref{eqn:UpdateRating} are the same as \eqref{eqn:BestResponse} and \eqref{eqn:RatingUpdateRule}, respectively. They are rewritten here for the convenience of reference. When determining the effort level in \eqref{eqn:UpdateEffort}, although the current payoff $u_i\left(m, \theta_i^t, d(\bm{\theta}^t), e_i, \bm{e}_{-i}^{t}\right)$ depends on the others' effort levels $\bm{e}_{-i}^t$, the current payoff can be separated into the benefit which depends only on the others' effort $\bm{e}_{-i}^t$, and the cost which depends only on agent $i$'s own effort $e_i$. Hence, when solving \eqref{eqn:UpdateEffort}, agent $i$ can treat the benefit as a constant, and consider only the cost, which depends on its own effort level and is known to itself.

The update of the parameter $\beta_i$ in \eqref{eqn:UpdateBelief} ensures that the conjectured future payoff equals to the current payoff, namely $f_i\left(\alpha_i,\beta_i^{t+1}, \theta_i^{t}, d(\bm{\theta}^t), e_i^{t}\right) = u_i\left(m,\theta_i^{t}, d(\bm{\theta}^{t}), \bm{e}^{t}\right)$.
When the system converges to a CE $\{\theta_i^*,e_i^*,\beta_i^*\}_{i \in \mathcal{N}}$, we will have $f_i\left(\alpha_i,\beta_i^{*}, \theta_i^{*}, d(\bm{\theta}^*), e_i^{*}\right) = u_i\left(m,\theta_i^{*}, d(\bm{\theta}^{*}), \bm{e}^{*}\right)$, which fulfills the third requirement of ``consistent conjectures'' in the definition of CE.

Next, we will provide the design guidelines on the matching rules, such that the above dynamics \eqref{eqn:UpdateEffort}--\eqref{eqn:UpdateBelief} always converge to a CE from any initial ratings. In fact, the design guideline is simple and intuitive: the matching rule should ensure that each agent's expected benefit is concave and increasing in its own rating.

\begin{definition}[Desirable Matching Rules]\label{definition:Rating-Fairness}
A matching rule $m$ is \emph{desirable}, if under any rating profile $\bm{\theta}$,
\begin{itemize}
\item each agent $i$'s (conjectured) expected benefit from the reviewing of its product, namely 
$$
\sum_{k_j \neq k_i} \left[ m_{k_i k_j}\left( d(\bm{\theta})_{k_i}, d(\bm{\theta})_{k_j} \right) \cdot b_i\left( d(\bm{\theta})_{k_j} \right) \right],
$$ 
is concave and increasing in its own rating $d(\bm{\theta})_{k_i}$;
\item each agent $i$'s expected number of products to review is positive and fixed, namely 
$$
\sum_{k_j \neq k_i} m_{k_j k_i}\left(d(\bm{\theta})_{k_j},d(\bm{\theta})_{k_i}\right) = M > 0.
$$
\end{itemize}
\end{definition}

The requirements of concavity and monotonicity are very reasonable. The expected benefit should be increasing in one's rating, such that one has incentives to exert high effort levels to increase its rating. In addition, if the expected benefit is concave in one's rating, since the marginal benefit is decreasing, one will not dramatically increase its effort level, which facilitates the convergence. The requirement of a fixed number of products to review ensures the fairness among the reviewers across time.

Despite the simplicity of the requirements for desirable matching rules, we are able to prove its convergence under proper rating update rules.

\begin{theorem}[Convergence]\label{theorem:Convergence}
Under any desirable matching rule, starting from any initial $\bm{\theta}^0$, there exists $\bar{\mu}>0$ such that under any small step size $\mu \in (0, \bar{\mu}]$ in the rating update rule, the system will converge to a CE through updates \eqref{eqn:UpdateEffort}--\eqref{eqn:UpdateBelief}.
\end{theorem}
\begin{IEEEproof}
See Appendix~\ref{proof:Convergence}.
\end{IEEEproof}

\begin{figure}
\centering
\includegraphics[width =3.6in]{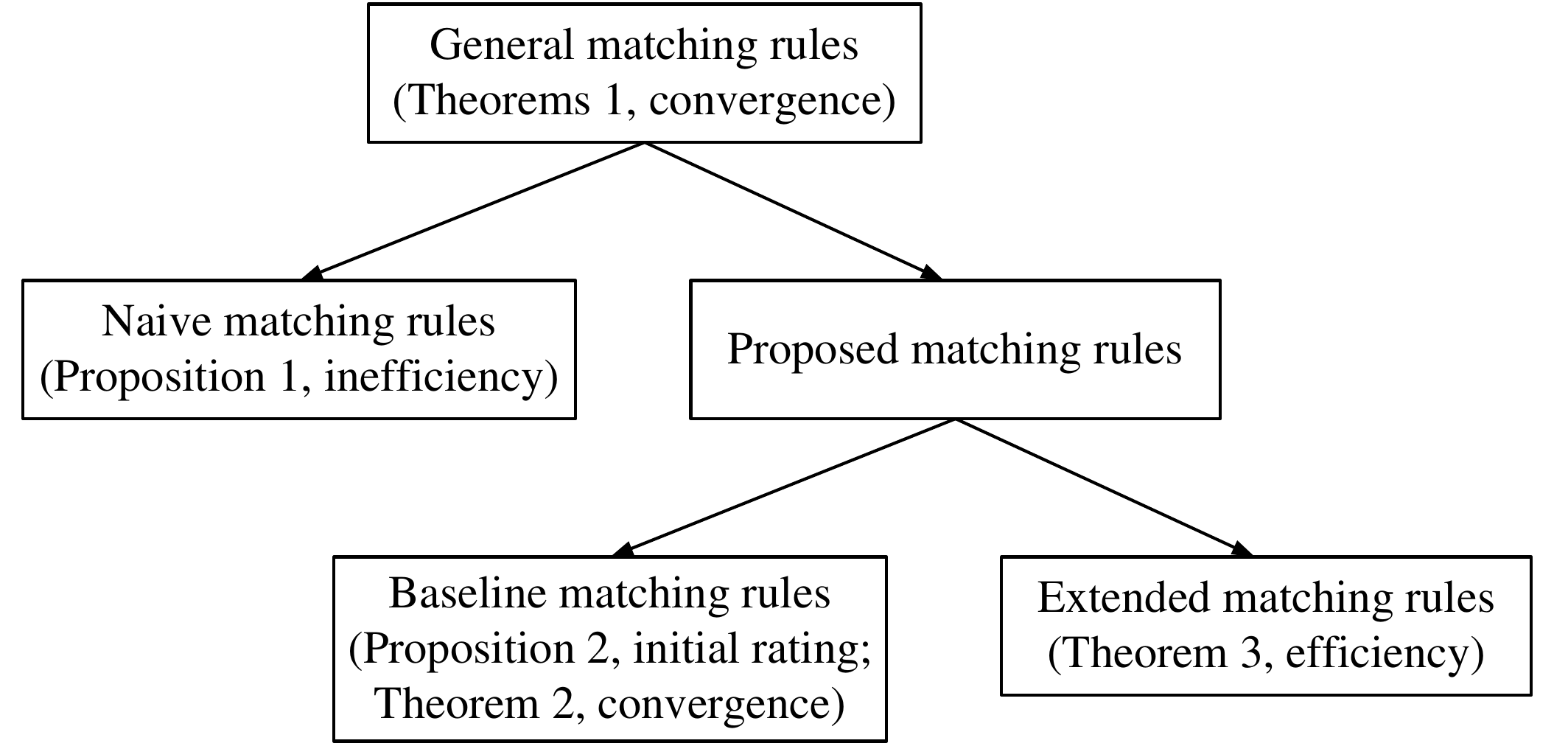}
\caption{Illustration of theoretical results.} \label{fig:IllustrationOfResults}
\end{figure}

We illustrate the theoretical results in this paper in Fig.~\ref{fig:IllustrationOfResults}.

\begin{remark}[Uniqueness]
Theorem~\ref{theorem:Convergence} proves that any desirable matching rule ensures the convergence to a CE. However, it is difficult to prove the convergence to a  particular CE. The reasons are that there is not a single unique CE and the asymptotically reached CE (through our matching and update rules) depends on the history of the system and thus the probabilistic matching assignments and the initial ratings. In our technical analysis we prove the convergence by showing that under the updates \eqref{eqn:UpdateEffort}--\eqref{eqn:UpdateBelief}, the difference (in terms of $\ell$-1 norm) between two consecutive rating profiles strictly decreases over time. However, since the best responses are different under different rating profiles, the mappings from the current rating profile to the next one are different over time. Hence, the contraction mapping theorem does not apply here. In fact, in Section~\ref{sec:design}, we will show that under several desirable matching rules, there are \emph{indeed multiple CEs}, and the system converges to different CEs under different initial ratings.
\end{remark}

\begin{remark}[Largest Step Size]
For the convergence, we require the step size $\mu$ in the rating update rule to be small enough. However, we would like the step size to be as large as possible, subject to convergence, for two reasons. First, a larger step size results in a fast convergence of the ratings to the true review quality. Second, perhaps less obviously, a larger step size provides higher incentives for agents to exert high effort levels. This is because with a larger step size, the influence of the current review quality is higher on the next rating. In Appendix~\ref{proof:Convergence}, we derive explicit upper bounds for the eligible step sizes. 
\end{remark}

\section{Design of Matching Rules}\label{sec:design}
The matching rule is the critical component of our design. In this section, we will first prove that existing matching rules are inefficient. Then we propose a baseline matching rule, and analyze the properties of this baseline matching rule in detail. Finally, we propose and study two extensions.

\subsection{Inefficiency of Existing Matching Rules}
We show the inefficiency of existing matching rules. In existing matching rules, the matching probabilities associated with agent $i$, $m_{k_i k_j}\left(d(\bm{\theta})_{k_i}, d(\bm{\theta})_{k_j}\right)$, do not depend on $i$'s ranking $k_i$ or its rating $d(\bm{\theta})_{k_i}$. This is true for existing matching rules in mechanism design \cite{Shapley}--\cite{RothSonmez}, because there is no notion of effort and hence no rating. It is also true for existing matching rules in social norms \cite{Kandori_SocialNorm}--\cite{XiaoMihaela_TEAC}, which are uniformly random.

\begin{proposition}
Under any matching rule that is independent of the rating of the agent whose product will be reviewed, namely $m_{k_i k_j}\left(d(\bm{\theta})_{k_i}, d(\bm{\theta})_{k_j}\right) = m_{k_i^\prime k_j}\left(d(\bm{\theta})_{k_i^\prime}, d(\bm{\theta})_{k_j}\right),~\forall k_i, k_i^\prime, k_j, \bm{\theta}$, there is a unique CE, in which $e_i^*=0$ and $\theta_i^*=q_i(e_i^*)=0$ for all $i$.
\end{proposition}
\begin{IEEEproof}
See Appendix II.
\end{IEEEproof}

The above proposition shows that the matching rule that does not depend on agents' ratings is the worst-case matching rule that results in ``free-riding'' by everyone. This underlines the importance of designing efficient matching rules that take into account the ratings of both the reviewers and the agents whose products are reviewed.

\subsection{Design of The Baseline Matching Rule}
The baseline matching rule works as follows:
\begin{enumerate}
\item For the agents with the same rating, match their products among themselves using any one-to-one mapping that does not match one's product to itself. 
\item For any agent $i$ with a distinct rating (i.e., no other agent has the same rating),
\begin{enumerate}
\item If it has the highest rating (i.e., $k_i=1$), match its product to a reviewer with the second highest rating with probability $1$.
\item If it has the lowest rating (i.e., $k_i=K$), match its product to a reviewer with the second lowest rating with probability $\frac{d(\bm{\theta})_K}{d(\bm{\theta})_{K-1}}$. Hence, its product gets no review with probability $1-\frac{d(\bm{\theta})_K}{d(\bm{\theta})_{K-1}}$.
\item If $1<k_i<K$, match its product to its two ``neighbors'' with the following probabilities (which sum up to $1$): 
\begin{eqnarray*}
\!\!\!\!\!\!\!\!\!\!\!\!&m_{k_i, k_i-1}(d(\bm{\theta})_{k_i},d(\bm{\theta})_{k_i-1}) = \frac{d(\bm{\theta})_{k_i}-d(\bm{\theta})_{k_i+1}}{d(\bm{\theta})_{k_i-1}-d(\bm{\theta})_{k_i+1}},
\end{eqnarray*}
and
\begin{eqnarray*}
\!\!\!\!\!\!\!\!\!\!\!\!&m_{k_i,k_i+1}(d(\bm{\theta})_{k_i},d(\bm{\theta})_{k_i+1}) = \frac{d(\bm{\theta})_{k_i-1}-d(\bm{\theta})_{k_i}}{d(\bm{\theta})_{k_i-1}-d(\bm{\theta})_{k_i+1}}.
\end{eqnarray*}
\end{enumerate}
\end{enumerate} 

\begin{figure}
\centering
\includegraphics[width =3.2in]{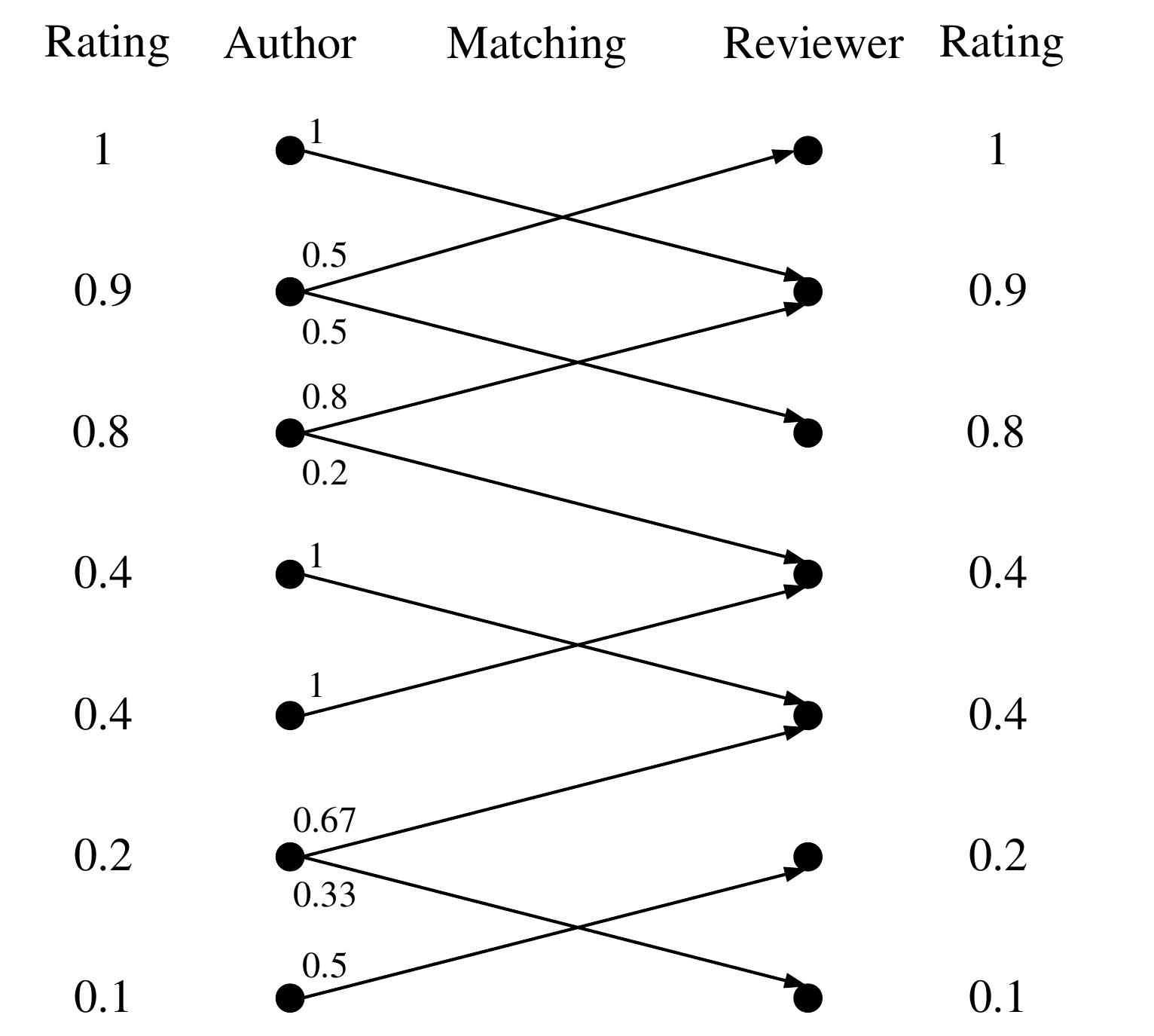}
\caption{Illustration of the baseline matching rule with 7 agents. The highest-rating agent's product is matched to the reviewer with the second highest rating. The lowest-rating agent's product is matched to the reviewer with the second lowest rating with probability $0.5$. The two agents with the same rating $0.4$ are matched to each other. The rest are matched to their two nearest neighbors with probabilities inversely proportional to the distances in ratings.} \label{fig:BaselineMatchingRule}
\end{figure}

The above matching rule is illustrated in Fig.~\ref{fig:BaselineMatchingRule}. Agents with the same rating are matched to each other. For an agent with a distinct rating, it matches its product with its two nearest ``neighbors" with probabilities that depend on how close its rating is to its neighbors' ratings. 

We propose this matching rule, because it has the following desirable properties:
\begin{itemize}
\item No agent will have to review more than 3 products. This is because any agent will at most review a product from an agent with the same rating (if there is any), and two products from its neighbors (if they have distinct ratings).
\item As we will prove later, this matching rule is a desirable matching rule as defined in Definition \ref{definition:Rating-Fairness}.
\end{itemize}

\subsubsection{The Choice of Initial Ratings}
In the considered system, it is important to choose the initial ratings correctly, because under different initial ratings, the system may converge to different CEs. Since the designer has no knowledge about the agents at the beginning, it is reasonable to assign the same initial rating to all the agents for fairness. In this case, the following proposition tells us that we should not make the initial rating too low.

\begin{proposition}\label{proposition:initial_rating}
There always exists a rating $\underline{\theta}$, such that any initial rating profile with the same rating $\theta^0 \leq \underline{\theta}$ for all agents is the equilibrium rating profile, and that each agent $i$ chooses an equilibrium effort level $e_i^*$ such that $q_i(e_i^*)=\theta^0$.
\end{proposition}

Proposition \ref{proposition:initial_rating} implies that we should choose a high enough initial rating. In particular, when the initial rating is too low, it is optimal to choose an effort level $e_i^*$ that satisfies $q_i(e_i^*)=\theta^0$. The key reason is that no agent has an incentive to reach a higher rating than the initial one, because in this case it will get a \emph{distinct} highest rating, and get the same benefit but a higher cost compared to choosing an effort level such that its rating remains the same as the initial rating. 
In other words, the initial rating determines the highest review quality produced by each agent. 

\subsubsection{Convergence}
It is useful to classify agents into types based on their cost, review quality, and benefit functions, etc. We define agents of a certain type as follows.
\begin{definition}[Types]\label{definition:Types}
The agents of the same \emph{type} have the same normalized marginal benefit to cost ratio, defined as $\frac{\delta_i \alpha_i q_i^\prime(\cdot)}{(1-\delta_i) c_i^\prime(\cdot)}$, the same review quality function $q_i(\cdot)$, and the same marginal benefit function $b_i^\prime(\cdot)$.
\end{definition}

\begin{definition}[Ordering of Capability]\label{definition:capability}
An agent $i$ is \emph{more capable} than an agent $j$, if
\begin{eqnarray*}
\frac{\delta_i \alpha_i q_i^\prime(e)}{(1-\delta_i) c_i^\prime(e)} &\geq& \frac{\delta_j \alpha_j q_j^\prime(e)}{(1-\delta_j) c_j^\prime(e)}, \forall e, \nonumber \\
q_i(e) &>& q_j(e), \forall e, \\
b_i^\prime(\theta) &\geq& b_j^\prime(\theta), \forall \theta. \nonumber
\end{eqnarray*}
\end{definition}

Definition~\ref{definition:Types} defines ``types'' of agents, in the sense that agents of the same type will always choose the same effort level and hence get the same rating. Definition~\ref{definition:capability} gives an ordering of agents in terms of their ``capability''. We will prove that a more capable agent indeed gets a higher rating.

In the rest of this section, we make the following assumption about the population size.

\begin{assumption}[Large Population]\label{assumption:Population}
There is more than one agent of each type.
\end{assumption}

Assumption~\ref{assumption:Population} is reasonable in practice, since the number of agents in peer review systems is indeed large. Given the same initial rating, the agents of the same type will choose the same best response effort level, and hence have the same rating. Assumption~\ref{assumption:Population} ensures that for each agent, there is always another agent with the same rating. According to Property 1) in the baseline matching rule, each agent will always have exactly one product to review all the time.

\begin{theorem}\label{theorem:convergence_baseline}
Suppose that the large population assumption (Assumption~\ref{assumption:Population}) holds. Then we have
\begin{itemize}
\item the baseline matching rule is a desirable matching rule; 
\item starting from any initial rating profile, there exists $\bar{\mu}>0$ such that under any small step size $\mu \in (0, \bar{\mu}]$ in the rating update rule, the system will converge to a CE through the best response dynamics \eqref{eqn:UpdateEffort}--\eqref{eqn:UpdateBelief};
\item if the agents have the same initial rating, at any point in the best response dynamics \eqref{eqn:UpdateEffort}--\eqref{eqn:UpdateBelief}, more capable agents will always have no lower ratings than less capable agents.
\end{itemize}
\end{theorem}
\begin{IEEEproof}
See Appendix II.
\end{IEEEproof}

Theorem \ref{theorem:convergence_baseline} ensures the convergence of the best response dynamics to a CE. In fact, we can say something stronger about the best response dynamics. That is, a more capable agent never has lower ratings than a less capable agent \emph{at any point} in the best response dynamics. This means that the rating mechanism can distinguish the agents of different types, and rank them in the correct order. Note that more capable agents produce reviews of high enough quality (that result in higher ratings) in their self-interest, as a result of maximizing their own payoffs; they are not obliged to do so by the designer.

\begin{figure}
\centering
\includegraphics[width =3.2in]{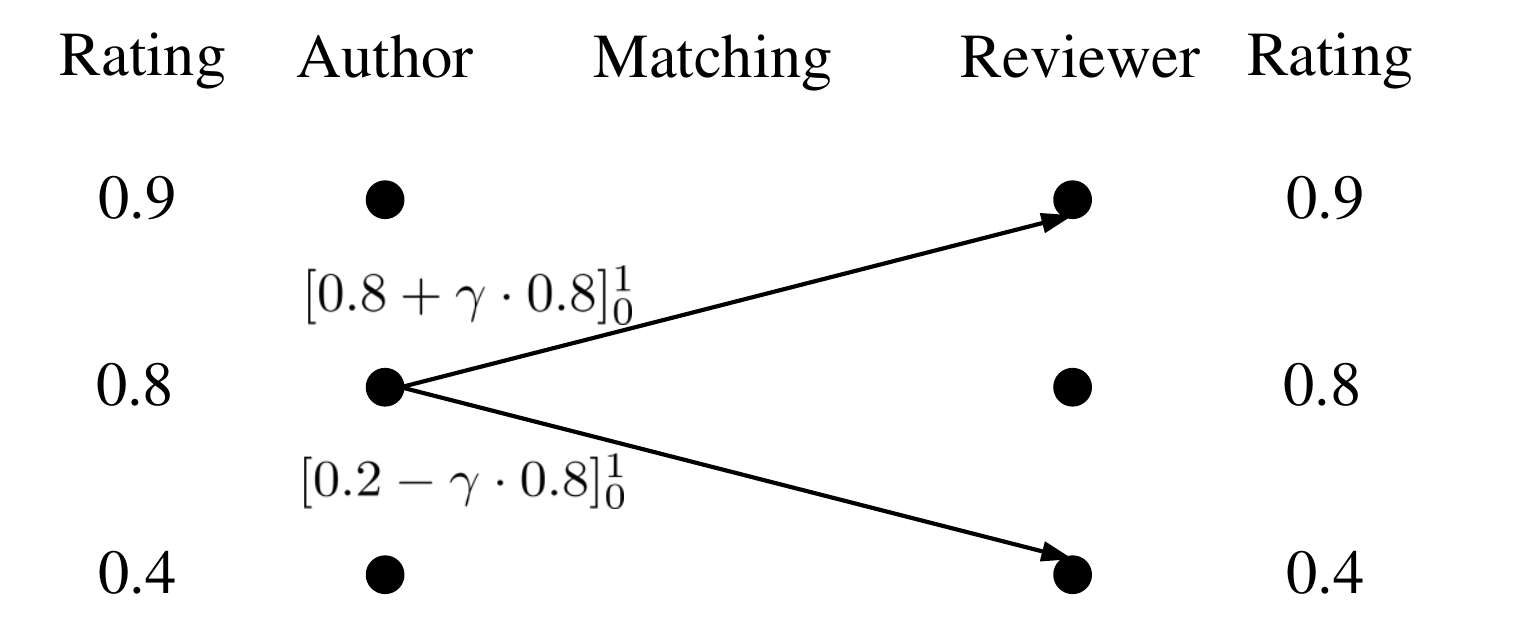}
\caption{An illustration of the first asymmetric extension of the baseline matching rule. We only show the matching of the agent with rating $0.8$, who is matched to its two nearest neighbors with different probabilities than in the baseline rule.} \label{fig:MatchingRule_Extension1}
\end{figure}

\subsection{Two Classes of Extended Matching Rules}
Previously, we have focused on the baseline matching rule. The baseline matching rule is able to incentivize the agents to exert high effort levels by increasing the benefit obtained by an agent when its rating increases. Now we extend the baseline rule in two different ways, both of which result in a class of matching rules that allow us to tune the reward and/or punishment provided by the matching rules.

In the first extension, we assign asymmetric probabilities for matching an agent with a distinct rating to its higher and lower neighbors. In particular, the {\em asymmetric matching rule} is parametrized by $\gamma$ such that any agent $i$ with a distinct rating and with $k_i \in [2, K-1]$ is matched to its neighbors with the following probabilities:
\begin{eqnarray*}
&\!\!\!\! m_{k_i, k_i-1}(d(\bm{\theta})_{k_i},d(\bm{\theta})_{k_i-1}) = \left[\frac{d(\bm{\theta})_{k_i}-d(\bm{\theta})_{k_i+1}}{d(\bm{\theta})_{k_i-1}-d(\bm{\theta})_{k_i+1}} + \gamma \cdot \theta_i\right]_0^1,
\end{eqnarray*} 
and
\begin{eqnarray*}
&\!\!\!\! m_{k_i,k_i+1}(d(\bm{\theta})_{k_i},d(\bm{\theta})_{k_i+1}) = \left[\frac{d(\bm{\theta})_{k_i-1}-d(\bm{\theta})_{k_i}}{d(\bm{\theta})_{k_i-1}-d(\bm{\theta})_{k_i+1}} - \gamma \cdot \theta_i\right]_0^1,
\end{eqnarray*}
where $[\cdot]_0^1 \triangleq \min\{ \max\{\cdot, 0\} , 1\}$.

We illustrate this asymmetric extension in Fig.~\ref{fig:MatchingRule_Extension1}.

We can see that when $\gamma>0$ ($\gamma<0$), the resulting matching rule rewards (punishes) the agent by increasing its probability of being matched to the higher-rating (lower-rating) neighbor. When $\gamma=0$, the asymmetric matching rule reduces to the baseline matching rule.

In the second extension, we allow an agent to be matched to a reviewer with even higher or even lower ratings than its two nearest neighbors. In particular, the matching rule is parametrized by $\gamma_r \in [0,1]$ and $\gamma_p \in [0,1]$. Then any agent $i$ with a distinct rating and with $k_i \in [3, K-2]$ is matched to its neighbors and neighbors of neighbors with the following probabilities:
\begin{eqnarray*}
m_{k_i, k_i-1}(d(\bm{\theta})_{k_i},d(\bm{\theta})_{k_i-1}) =& \frac{d(\bm{\theta})_{k_i}-d(\bm{\theta})_{k_i+1}}{d(\bm{\theta})_{k_i-1}-d(\bm{\theta})_{k_i+1}} \cdot (1-\gamma_r),\\
m_{k_i, k_i-2}(d(\bm{\theta})_{k_i},d(\bm{\theta})_{k_i-2}) =& \frac{d(\bm{\theta})_{k_i}-d(\bm{\theta})_{k_i+1}}{d(\bm{\theta})_{k_i-1}-d(\bm{\theta})_{k_i+1}} \cdot \gamma_r,
\end{eqnarray*}
and
\begin{eqnarray*}
m_{k_i,k_i+1}(d(\bm{\theta})_{k_i},d(\bm{\theta})_{k_i+1}) =& \frac{d(\bm{\theta})_{k_i-1}-d(\bm{\theta})_{k_i}}{d(\bm{\theta})_{k_i-1}-d(\bm{\theta})_{k_i+1}} \cdot (1-\gamma_p), \\
m_{k_i,k_i+2}(d(\bm{\theta})_{k_i},d(\bm{\theta})_{k_i+2}) =& \frac{d(\bm{\theta})_{k_i-1}-d(\bm{\theta})_{k_i}}{d(\bm{\theta})_{k_i-1}-d(\bm{\theta})_{k_i+1}} \cdot \gamma_p.
\end{eqnarray*}

\begin{figure}
\centering
\includegraphics[width =3.2in]{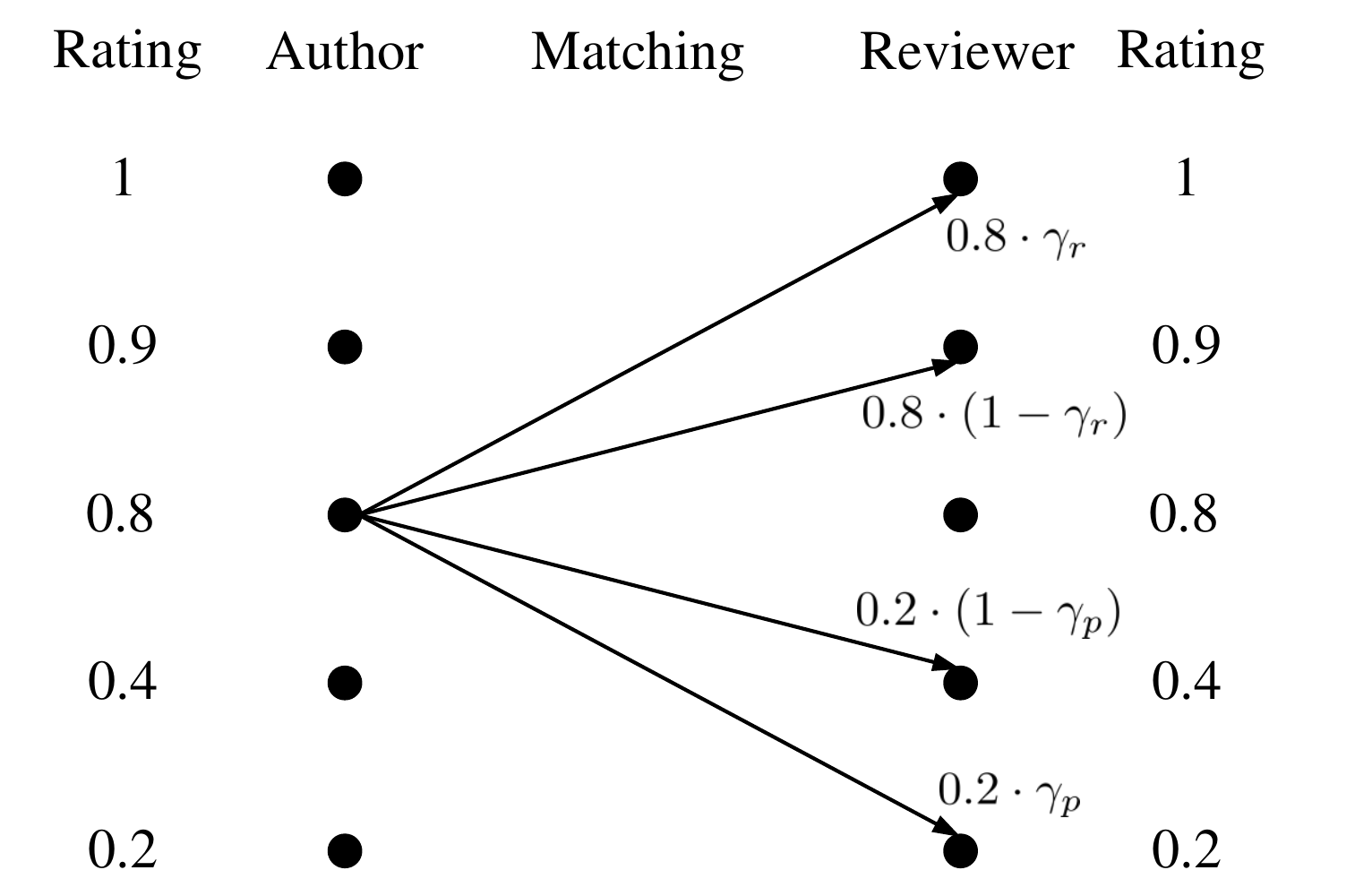}
\caption{An illustration of the second long-range extension of the baseline matching rule. We show the matching of only one agent with rating $0.8$, who is matched to its four neighbors, instead of two nearest neighbors as in the baseline rule.} \label{fig:MatchingRule_Extension2}
\end{figure}

We refer to this extension as {\em long-range matching rule}; see Fig.~\ref{fig:MatchingRule_Extension2} for an illustration.

We can see that the parameters $\gamma_r$ and $\gamma_p$ reflect to what extent the agents are rewarded and punished, respectively. When $\gamma_r=\gamma_p=0$, the matching rule reduces to the baseline rule. When $\gamma_r=1$ ($\gamma_p=1$), the agent is rewarded (punished) by being matched to a reviewer with the next higher (lower) rating.

\begin{figure}
\centering
\includegraphics[width =3.2in]{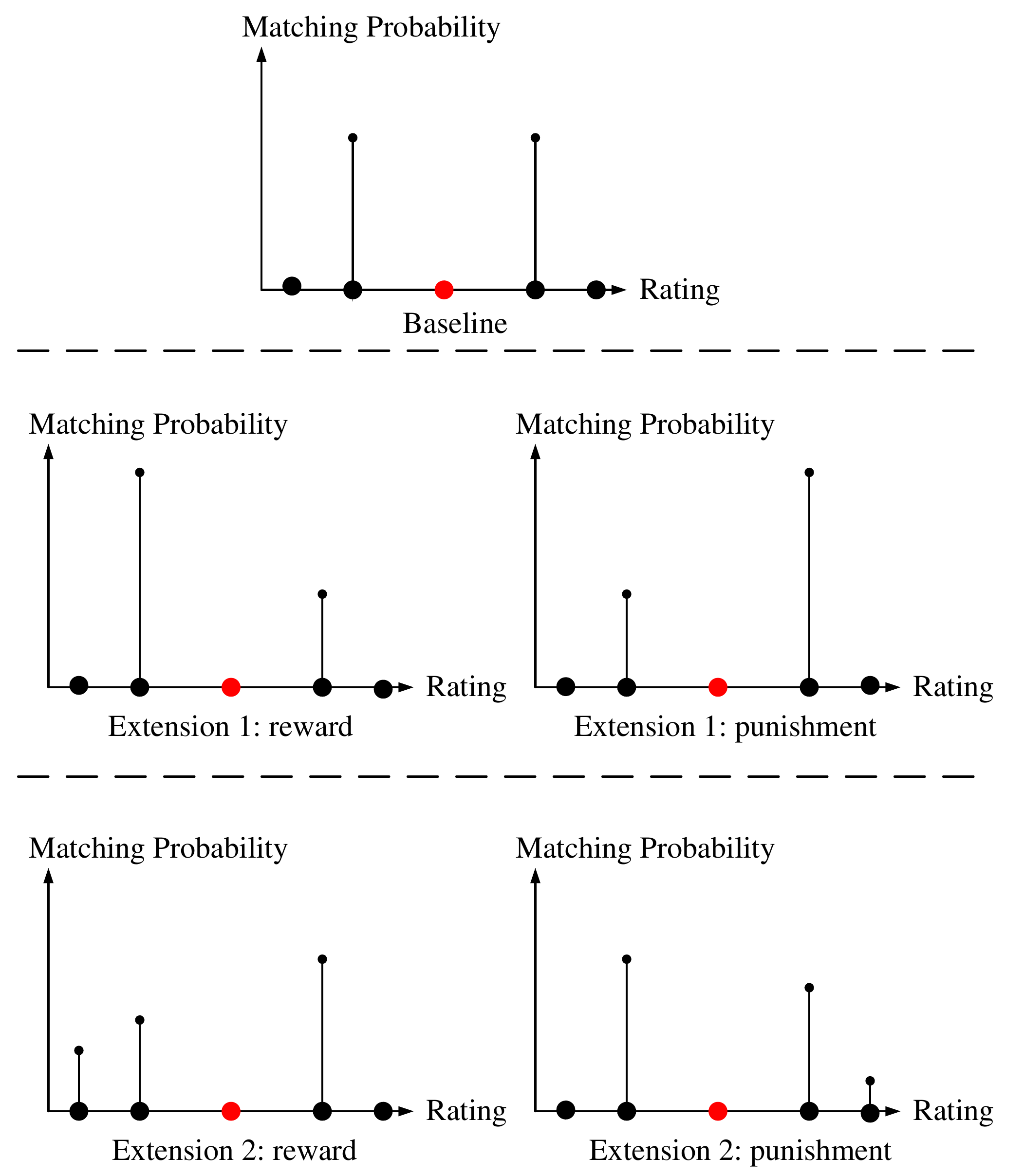}
\caption{Comparison of the baseline matching rule and its two extensions. We illustrate how an agent with a certain rating (the red dot) is matched. In the baseline rule, it is matched to its nearest neighbors with equal probability. In the first asymmetric extension, although the distances between the ratings remain the same, the matching probabilities are slightly changed. In the second long-range extension, it can be matched to neighbors with even higher or lower ratings.} \label{fig:MatchingRuleComparison}
\end{figure}

We summarize the key differences among the baseline matching rule and its extensions in Fig.~\ref{fig:MatchingRuleComparison}.

It is interesting to ask under each class of extended matching rules, which matching rule is optimal in terms of the equilibrium review quality? We first define the notion that one matching rule is ``better'' than the other.
\begin{definition}
We say that a matching rule $m^\prime$ is ``better'' than another matching rule $m$, if for any equilibrium rating profile $\bm{\theta}^*$ under $m$, we can find an equilibrium rating profile $\bm{\theta}^{*\prime}$ under $m^\prime$ that satisfies $\bm{\theta}^{*\prime}>\bm{\theta}^*$.
\end{definition}

The following theorem tells us how to design an extended matching rule that is better than the baseline rule.

\begin{theorem}\label{theorem:extensions}
Suppose that the large population assumption (Assumption~\ref{assumption:Population}) holds. Then we have:
\begin{itemize}
\item
In the first asymetric extension, there exists a $\gamma>0$ (i.e., extra reward) under which the asymmetric matching rule is strictly better than the baseline matching rule.
\item In the second long-range extension, there exists $\gamma_r=0$ and $\gamma_p>0$ (i.e., extra punishment) under which the long-range matching rule is strictly better than the baseline matching rule.
\end{itemize}
\end{theorem}

Theorem \ref{theorem:extensions} tells us if we reward or punish by assigning higher or lower probabilities of being matched to the higher-rating neighbor, it is beneficial to reward. On the contrary, if we reward or punish by creating the possibility of being assigned to the next higher- or lower-rating neighbors, it is beneficial to punish. Note that we can get the benefit only when we set the correct parameters in the two extended matching rules. The technical reason is that we want to increase the marginal expected benefit when an agent's rating is changed, in order to give more incentive for them to exert high effort levels. The main message delivered by our result is that we should carefully design the matching rule based on the way we reward and punish.

\section{Illustrative Results}\label{sec:simulation}
We consider a system with 10 types of agents. There are 100 agents of each type. All the agents have the same patience $\delta_i=0.8$, the same cost function $c_i(e_i) = e_i^2$, and the same benefit function $b_i(\theta) = -\theta^2+2\theta$. Different types of agents have different quality functions $q_i(e_i) = p_i\cdot e_i$ with $p_i=0.2, 0.4, 0.6, \ldots, 2.0$. They also have different $\alpha_i = 0.2, 0.4, 0.6, \ldots, 2.0$ in the conjecture functions.

\subsection{Impact of Step Sizes in Rating Update}
\begin{figure}
\centering
\includegraphics[width =2.6in]{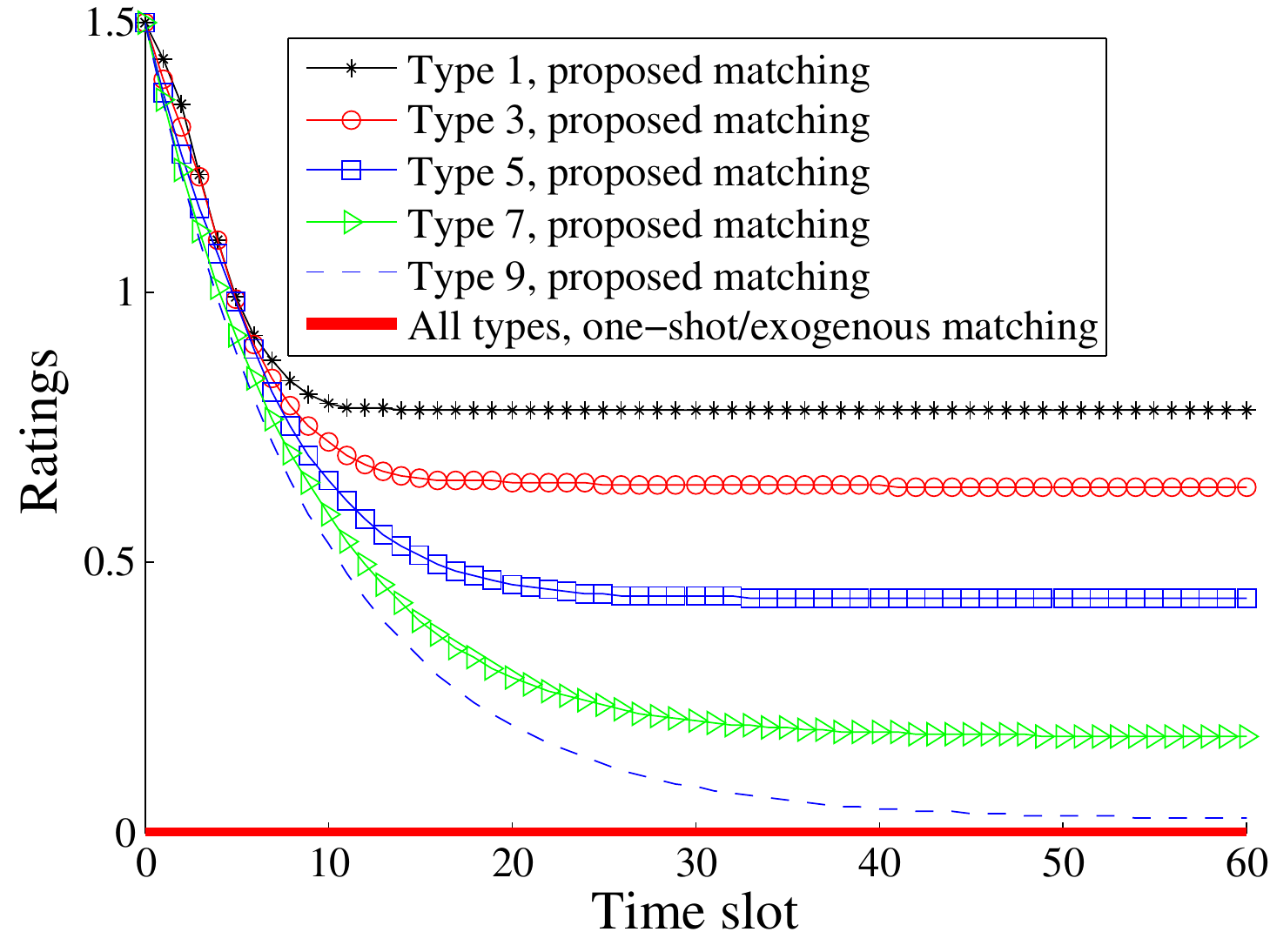}
\caption{Convergence under step size $\mu=0.1$.} \label{fig:LowStepSize}
\end{figure}

\begin{figure}
\centering
\includegraphics[width =2.6in]{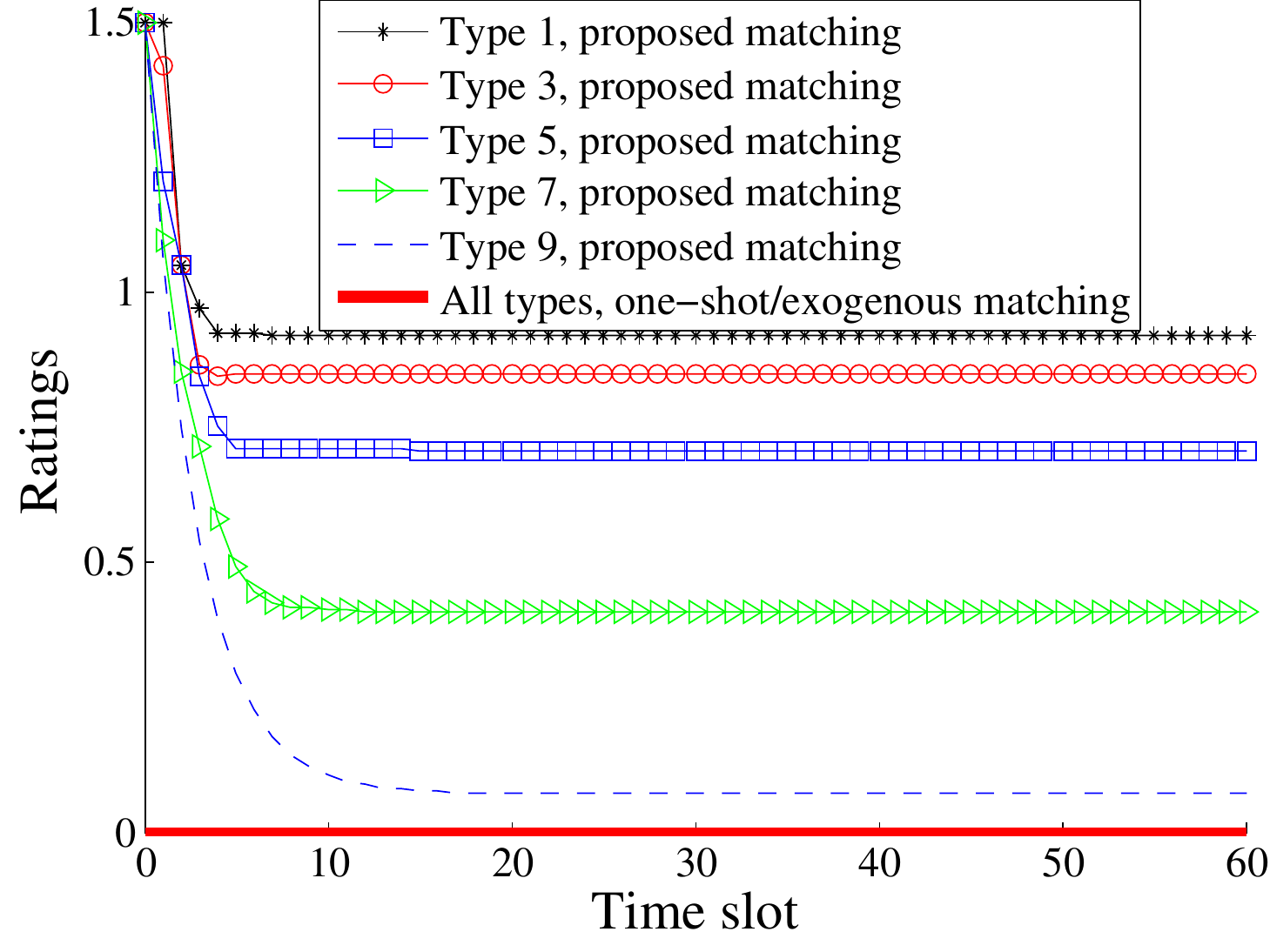}
\caption{Convergence under step size $\mu=0.3$.} \label{fig:HighStepSize}
\end{figure}

We show the best response dynamics under step sizes $\mu=0.1$ and $\mu=0.3$ in Fig.~\ref{fig:LowStepSize} and Fig.~\ref{fig:HighStepSize}, respectively. Note that we only show the ratings of agents of types $1,3,5,7,9$. We first observe that under one-shot or exogenous matching rules, the reviewers exert 0 efforts and get 0 ratings all the time. Hence, the proposed endogenous matching greatly improves the performance of the system.

Second, we can see that under a larger step size, the agents' equilibrium ratings are higher, indicating higher equilibrium effort levels and higher equilibrium review quality. This is consistent with our intuition in Remark 2. Note that, when $\mu=0.5$, the best response dynamics do not converge (type-1 agents' ratings are oscillating). This stresses the importance of choosing the step size: We want to choose a step size as high as possible for better review quality, subject to the constraint that the best response dynamics converge.

\subsection{Different Matching Rules}
We compare the sum review quality and the social welfare (i.e., the total benefit minus cost) at the equilibrium under different matching rules.

In Table~\ref{table:Extension1}, we evaluate the first asymmetric extension of matching rules under different parameters $\gamma$. We can see that in our setting, the optimal $\gamma$ should be 0.1, which results in the highest sum review quality. This is consistent with our theoretical results: we can find a rewarding matching rule that outperforms the baseline rule. It is worth mentioning that the matching rule that maximizes the sum review quality may not be the one that maximizes the social welfare. This is reasonable because higher review quality also results in higher cost. In fact, in terms of social welfare, the optimal $\gamma$ is -0.05, which results in lower review quality and thus lower cost. How the review quality and the social welfare are aligned depends on the benefit and cost functions.

In Table~\ref{table:Extension2}, we evaluate the second long-range extension of matching rules under different parameters $\gamma_r$ and $\gamma_p$. We can see that the optimal sum review quality is achieved when $\gamma_r=0$ and $\gamma_p=1$, which is a matching rule that punishes to the most severe extent. The threat of being matched to an even lower-rating reviewer provides more incentive for agents to exert high effort. Again, such a matching rule does not result in the optimal social welfare. The optimal social welfare is achieved when $\gamma_r=0.5$ and $\gamma_p=0.5$, where the agents are also rewarded.

\section{Conclusion}\label{sec:conclusion}
In this work, we proposed the first rating and repeated endogenous matching mechanisms to address the adverse selection and moral hazard problems simultaneously in peer review. Our proposed rating and matching mechanisms are easy to implement, require no knowledge of agents' private information, and ensure the convergence to an equilibrium in which the agents get their review quality revealed by their ratings and are incentivized to produce high-quality reviews. We thoroughly studied the design of matching rules, in terms of the initial ratings, the requirements for convergence, and the equilibrium ratings and review quality. We also studied extensions to different classes of matching rules, and proved the optimality of different rewarding/punishing mechanisms under different matching rules.

In future work, we will investigate the effect of inaccurate and possibly biased reports about the review quality, as well as more detailed modeling frameworks including multiple agents and reviewers associated with each product. An intriguing problem is the quest for the \emph{best} matching rule maximizing the review quality.

\begin{table}\scriptsize
\renewcommand{\arraystretch}{1.1}
\caption{Equilibrium review quality and social welfare under the first asymmetric extension of matching rules.} \label{table:Extension1} \centering
\begin{tabular}{|c|c|c|c|c|c|c|c|}
\hline
$\gamma$ & -0.2 & -0.1 & -0.05 & 0 & 0.05 & 0.1 & 0.2 \\
\hline
Sum review quality & 0.64 & 0.91 & 0.96 & 1.29 & 1.28 & \textbf{1.36} & 1.28 \\
\hline
Social welfare & 1.37 & 1.58 & \textbf{1.59} & 1.44 & 1.45 & 1.46 & 1.55 \\
\hline
\end{tabular}
\end{table}

\begin{table}\scriptsize
\renewcommand{\arraystretch}{1.1}
\caption{Equilibrium review quality and social welfare under the second long-range extension of matching rules.} \label{table:Extension2} \centering
\begin{tabular}{|c|c|c|c|c|c|c|}
\hline
$(\gamma_r,\gamma_p)$ & (0, 0) & (0, .5) & (0, 1) & (.5, 0) & (.5, .5) & (.5, 1) \\
\hline
Sum review quality & 1.29 & 1.31 & \textbf{1.40} & 1.11 & 1.28 & 1.33 \\
\hline
Social welfare & 1.44 & 1.41 & 1.35 & 1.27 & \textbf{1.57} & 1.43 \\
\hline
\end{tabular}
\end{table}

\appendices

\section{Properties of best response}\label{Appendix: Properties of best response}

For ease of reference, we recall agent $i$'s best response  $e_i^t$ in time slot $t$ here:
\begin{eqnarray}
e_i^t = \arg \max_{e_i \in \left[0, e_i^{\rm max} \right] } & & (1-\delta_i) \cdot u_i(m, \theta_i^t, d(\bm{\theta}^t), e_i, \bm{e}_{-i}^t) \nonumber\\
&+& \delta \cdot f_i(\alpha_i,\beta_i^t, \theta_i^t, d(\bm{\theta}^t), e_i).
\label{eq: best response}
\end{eqnarray}
We denote the best response function by $B_i: \bm{\theta^t} \mapsto e_i^t$. We need the following lemma summarizing the properties of the best response for future reference.

\begin{lemma}\label{Lemma: Properties of best response}
Consider the current best response as defined in \eqref{eq: best response}. The following statements hold:
\begin{enumerate}

	\item The best response is unique for every $t \geq 0$;

	\item  The best response function is a continuous and almost everywhere differentiable function (uniformly in $t$) of $\theta^t$;
	
	\item For any $t \geq 0$, the best response function satisfies
	$
	\left| B_i(\bm{\theta}^{t+1})-B_i(\bm{\theta}^t) \right| \leq L_i(\mu) \cdot \left\| \bm{\theta}^{t+1}-\bm{\theta}^t \right\|_1,
	$
	where $L_i(\mu)$ is a continuous function of the step size $\mu$ that converges to zero as $\mu \downarrow 0$.

\end{enumerate}

\end{lemma}

\begin{IEEEproof}
The current payoff $u_i(m, \theta_i^t, d(\bm{\theta}^t), e_i, \bm{e}_{-i}^t)$ in the current best response \eqref{eq: best response} has two terms:
$$
\textstyle\sum_{k_j \neq k_i} m_{k_i k_j}\left(\theta_i^t,d(\bm{\theta}^t)_{k_j}\right) \cdot b_i\left(q_j(e_j^t)\right)
- M \cdot c_i(e_i),
$$
where the first term, namely the expected current benefit, does not depend on agent $i$'s own effort.
Similarly, the conjecture function $f_i(\alpha_i,\beta_i^t, \theta_i^t, d(\bm{\theta}^t), e_i) =
\alpha_i  \cdot \bar{b}_i\left(\theta_i^t, d(\bm{\theta}^t), e_i^t\right) + \beta_i^t$, 
where the second term $\beta_i$ does not depend on agent $i$'s own effort.  
Hence, we can simplify the best response as
\begin{eqnarray}\label{eqn:best_response}
e_i^t = \arg \max_{e_i \in [0, e_i^{\rm max}]} &&
- (1-\delta_i) M c_i(e_i) \\
&+& \delta_i \alpha_i   \bar{b}_i\left(\theta_i^t, d(\bm{\theta}^t), e_i^t\right). \nonumber
\end{eqnarray}

Since the cost $c_i(e_i)$ is strictly convex in $e_i$, since the grading quality $q_i(e_i)$ is concave in $e_i$, and since the conjectured benefit is concave and increasing in $(1-\mu) \theta_i^t + \mu q_i(e_i)$, the objective function in  \eqref{eqn:best_response} is strictly concave in $e_i$. Thus, the current best response is unique for each $t \geq 0$.

With regards to statement 2): due to strict concavity, the objective function in \eqref{eqn:best_response} is almost everywhere differentiable with respect to $e_i$ 
and its derivative is equal almost everywhere to a decreasing function of $e_i$ \cite[Exercise 1.6.42]{Tao}. 
Due to the Monotone Differentiation Theorem \cite[Theorem 1.6.25]{Tao}, the derivative itself is almost everywhere differentiable with respect to $e_i$ as well. 
We concluded that the objective function in \eqref{eqn:best_response} is almost everywhere twice differentiable with respect to $e_i$. 
As a consequence, the unique best response $e_i^t$ is almost everywhere differentiable with respect to $\bm{\theta^t}$ due to the Implicit Function Theorem \cite[Ch. 5, Theorem 2.1]{delaFuente}. 
In addition, since the objective function is continuous (any concave function is continuous \cite[Ch. 6, Theorem 2.14]{delaFuente}), and the feasible set does not depend on $\bm{\theta^t}$, the unique best response $e_i$ is continuous in $\bm{\theta^t}$ according to Berge's Theorem of Maximum \cite[Ch. 7, Theorem 2.1]{delaFuente}. In summary, the best response $e_i$ is continuous and almost everywhere differentiable of $\bm{\theta^t}$. Notice that all of these properties hold uniformly in $t$. 

With regards to statement 3):
Instead of studying $B_i$ directly, we study a closely related function $\hat{B}_i: \bm{\theta} \mapsto \hat{e}_i^t$ defined as
\begin{multline*}
\hat{e}_i^t = \arg \max_{e_i \in \mathbb{R}}
- (1-\delta_i) M c_i(e_i) +
\delta_i \alpha_i  \bar{b}_i\left(\theta_i^t, d(\bm{\theta}^t), e_i^t\right).
\end{multline*}
Note that the above optimization problem is almost the same as \eqref{eqn:best_response}, except that the feasible set is $e_i \in \mathbb{R}$ instead of $e_i \in [0, e_i^{\rm max}]$. In other words, the new optimization problem is unconstrained. Following the same logic, we know that $\hat{B}_i$ is continuous and almost everywhere differentiable. Moreover, since the optimization problem is  unconstrained,
we do not need to worry about the boundary of the feasible set. Hence, $\hat{e}_i^t$ is the unique solution to the first-order condition (\emph{whenever the derivative exists}):

\begin{eqnarray}\label{eqn:FOC}
0 &=& -(1-\delta_i) M c_i^\prime(\hat{e}_i^t)  \\ 
& + & \textstyle\delta_i \alpha_i \mu q_i^\prime(\hat{e}_i^t) \cdot \sum_{k_j \neq k_i^+} \Big[ \frac{\partial m_{k_i^+ k_j}(\theta_i^{t+1}, d(\theta_i^{t+1},\bm{\theta}_i^t)_{k_j})}{\partial \theta_i^{t+1}} \Big. \nonumber \\
& & ~~~~~~~~~~~~~~~~~~~~~~~~~~~~~~~~ \Big. \cdot b_i(d(\theta_i^{t+1},\bm{\theta}_i^t)_{k_j}) \Big], \nonumber
\end{eqnarray}
where $\theta_i^{t+1} = (1-\mu) \theta_i^t + \mu q_i(\hat{e}_i^t)$. Note that when $k_j \neq k_i^+$, $b_i(d(\theta_i^{t+1},\bm{\theta}_i^t)_{k_j}) = b_i(\theta_j^t)$ does not depend on $\theta_i^{t+1}$.
The best response function $B_i$ is related to $\hat{B}_i$ in the following way: 
$$
B_i(\bm{\theta}^t) = \left[ \hat{B}_i(\bm{\theta}^t) \right]_0^{e_i^{\rm max}},
$$
where $[\cdot]_0^{e_i^{\rm max}} = \min\left\{\max\left\{\cdot, 0\right\}, e_i^{\rm max} \right\}$ is the projection of a real number on the interval $[0, e_i^{\rm max}]$.
This is because at any $\bm{\theta}^t$, due to strict concavity of the objective function, the first-order derivative is decreasing and hence must be positive for all $e_i < \hat{B}_i(\bm{\theta}^t)$. In other words, the objective function is increasing in $e_i$ for all $e_i < \hat{B}_i(\bm{\theta}^t)$. If $\hat{B}_i(\bm{\theta}^t) > e_i^{\rm max}$, then the objective function is increasing in $e_i$ for all $e_i \leq e_i^{\rm max}$.
In this case, the maximum of the objective function in \eqref{eqn:best_response} is taken at $e_i^t = e_i^{\rm max}$. Similarly, $B_i(\bm{\theta}^t) = 0$ if $\hat{B_i}(\bm{\theta}^t) \leq 0$.

\begin{figure*}[!t]
\normalsize \setcounter{tempeqncnt}{\value{equation}} \setcounter{equation}{\value{equation}}
\begin{align}
\label{eqn:derivative_j} 
\frac{\partial \hat{B}_i(\bm{\theta}^t)}{\partial \theta_j^t} = 
\frac{
\mu \cdot \left\{ 
\delta_i \alpha_i  q_i^\prime(\hat{e}_i^t) 
\left[ 
\frac{\partial^2 m_{k_i^+ k_j}( \theta_i^{t+1}, d_{i,k_j}^+ )}{\partial \theta_i^{t+1} \partial \theta_j^t} b_i(\theta_j^t)
+ \frac{\partial m_{k_i^+ k_j}( \theta_i^{t+1}, d_{i,k_j}^+ )}{\partial \theta_i^{t+1}} b_i^\prime(\theta_j^t)
\right] 
\right\}
}
{
(1-\delta_i) M c_i^{\prime\prime}(\hat{e}_i) 
- \mu \delta_i \alpha_i \left\{ 
q_i^{\prime\prime}(\hat{e}_i^t) {\displaystyle\sum_{k \neq k_i^+}}  \left[ \frac{\partial m_{k_i^+ k}(\theta_i^{t+1}, d_{i,k}^+)}{\partial \theta_i^{t+1}} b_i(d_{i,k}^+) \right]
+ \mu \left[ q_i^{\prime}(\hat{e}_i^t) \right]^2 \frac{\partial^2 m_{k_i^+ k_j}( \theta_i^{t+1}, d_{i,k_j}^+ )}{\partial \left[\theta_i^{t+1}\right]^2} b_i(\theta^t_j) \right\}
}
\end{align}
\hrulefill \vspace*{4pt} 
\begin{align}
\label{eqn:derivative_i} 
\frac{\partial \hat{B}_i(\bm{\theta}^t)}{\partial \theta_i^t} = 
\frac{
\mu \cdot \left[ 
(1-\mu) \delta_i \alpha_i  q_i^\prime(\hat{e}_i^t) \sum_{k \neq k_i^+}
\frac{\partial^2 m_{k_i^+ k}( \theta_i^{t+1}, d_{i,k}^+ )}{\partial \left[ \theta_i^{t+1} \right]^2} b_i( d_{i,k}^+ )
\right]
}
{
(1-\delta_i) M c_i^{\prime\prime}(\hat{e}_i) 
- \mu \delta_i \alpha_i  
\sum_{k \neq k_i^+}  \left[ q_i^{\prime\prime}(\hat{e}_i^t) \frac{\partial m_{k_i^+ k}(\theta_i^{t+1}, d_{i,k}^+)}{\partial \theta_i^{t+1}} b_i(d_{i,k}^+) 
+ \mu \left[ q_i^{\prime}(\hat{e}_i^t) \right]^2 \frac{\partial^2 m_{k_i^+ k}( \theta_i^{t+1}, d_{i,k}^+ )}{\partial \left[\theta_i^{t+1}\right]^2} b_i( d_{i,k}^+ ) \right]
}
\end{align}
\hrulefill \vspace*{4pt} 
\end{figure*}

Now we can study $\hat{B}_i(\bm{\theta}^t)$. In particular, we want to know how $\hat{e}_i^t$ changes when $\bm{\theta}^t$ changes. To this end, we apply the Implicit Function Theorem to \eqref{eqn:FOC} by taking the derivative of the right-hand side of \eqref{eqn:FOC} with respect to $\theta_j^t$ and keeping in mind that $\hat{e}_i^t=\hat{B}_i(\bm{\theta}^t)$ is a function of $\theta_j^t$. Then we obtain $\frac{\partial \hat{B}_i(\bm{\theta}^t)}{\partial \theta_j^t}$ for $j \neq i$ in \eqref{eqn:derivative_j} and $\frac{\partial \hat{B}_i(\bm{\theta}^t)}{\partial \theta_i^t}$ in \eqref{eqn:derivative_i} shown at the top of the next page, where $d_i^+ \triangleq d\left(\theta_i^{t+1}, \bm{\theta}_{-i}^t\right)$ and $d_{i,k_j}^+ \triangleq d\left(\theta_i^{t+1}, \bm{\theta}_{-i}^t\right)_{k_j}=\theta_j^t$ for $k_j \neq k_i$. 

Based on the above derivations, we prove the following important property about the best response function $B_i$.

First, we have:
\begin{eqnarray*}
\left| B_i(\bm{\theta}^{t+1})-B_i(\bm{\theta}^t) \right|
&=& \left| \left[ \hat{B}_i(\bm{\theta}^{t+1}) \right]_0^{e_i^{\rm max}} - \left[ \hat{B}_i(\bm{\theta}^t) \right]_0^{e_i^{\rm max}} \right| \\
&\leq& \left| \hat{B}_i(\bm{\theta}^{t+1}) - \hat{B}_i(\bm{\theta}^t) \right|,
\end{eqnarray*}
where the inequality holds because the difference between the projections of two numbers on the interval is no larger than the difference between the two numbers. 

Next, we write $\hat{B}_i(\bm{\theta}^{t+1}) - \hat{B}_i(\bm{\theta}^t)$ as a ``telescoping sum'' of $N$ terms as follows:
\begin{eqnarray*}
& & \hat{B}_i(\bm{\theta}^{t+1}) - \hat{B}_i(\bm{\theta}^t) \\
&=& \textstyle\sum_{j=1}^N \left[ \hat{B}_i\left( \theta_1^{t+1},\ldots,\theta_{j-1}^{t+1},\theta_j^{t+1},\theta_{j+1}^t,\ldots,\theta_N^t \right) \right. \\
& & ~~~~~~ \textstyle- \left. \hat{B}_i\left( \theta_1^{t+1},\ldots,\theta_{j-1}^{t+1},\theta_j^t,\theta_{j+1}^t,\ldots,\theta_N^t \right) \right],
\end{eqnarray*}
where the $j$th term is the best response under the rating profile where the first $j$ agents have ratings in period $t+1$, minus the best response under the rating profile where the first $j-1$ agents have ratings in period $t+1$. As a result, we have:
\begin{eqnarray*}
& & \left| \hat{B}_i(\bm{\theta}^{t+1}) - \hat{B}_i(\bm{\theta}^t) \right| \\
&=& \textstyle\sum_{j=1}^N \left| \hat{B}_i\left( \theta_1^{t+1},\ldots,\theta_{j-1}^{t+1},\theta_j^{t+1},\theta_{j+1}^t,\ldots,\theta_N^t \right) \right. \\
& & ~~~~~~~~ \textstyle- \left. \hat{B}_i\left( \theta_1^{t+1},\ldots,\theta_{j-1}^{t+1},\theta_j^t,\theta_{j+1}^t,\ldots,\theta_N^t \right) \right| \\
&=& \sum_{j=1}^N \left| \int_{\theta_j^t}^{\theta_j^{t+1}} \!\!\! \frac{\partial \hat{B}_i\left( \theta_1^{t+1},\ldots,\theta_{j-1}^{t+1},\theta_j,\theta_{j+1}^t,\ldots,\theta_N^t \right)}{\partial \theta_j} d\theta_j \right| \\
&\leq& \textstyle\sum_{j=1}^N \left|\theta_j^{t+1}-\theta_j^t\right| \cdot \\
& & \mathrm{ess~sup}_{\theta_j \in [\theta_j^t, \theta_j^{t+1}]} \textstyle\left| \frac{\partial \hat{B}_i\left( \theta_1^{t+1},\ldots,\theta_{j-1}^{t+1},\theta_j,\theta_{j+1}^t,\ldots,\theta_N^t \right)}{\partial \theta_j} \right|,
\end{eqnarray*}
where we take the essential supremum, since the derivative may not exist for all $\theta_j \in [\theta_j^t, \theta_j^{t+1}]$, but exists almost everywhere except on a subset of measure zero.

We define
$$
L_{ij}(\mu) \triangleq \mathrm{ess~sup}_{\bm{\theta}} \frac{\partial \hat{B}_i(\bm{\theta})}{\partial \theta_i}(\mu),~\mathrm{subject~to}~\frac{\partial \hat{B}_i(\bm{\theta})}{\partial \theta_i}~\mathrm{exists},
$$
where we write $\frac{\partial \hat{B}_i(\bm{\theta})}{\partial \theta_i}(\mu)$ to emphasize the fact that $\frac{\partial \hat{B}_i(\bm{\theta})}{\partial \theta_i}$ is a function of the step size $\mu$. We further define
$$
\textstyle L_i(\mu) \triangleq \max_{j \in \mathcal{N}} L_{ij}(\mu).
$$

We can see from \eqref{eqn:derivative_j} and \eqref{eqn:derivative_i} that $\frac{\partial \hat{B}_i(\bm{\theta})}{\partial \theta_i}(\mu)$ is continuous in $\mu$ and is $0$ when $\mu=0$. Hence, due to Berge's Theorem of Maximum \cite[Ch. 7, Theorem 2.1]{delaFuente}, $L_{ij}(\mu)$ is continuous in $\mu$, and $L_{ij}(0)=0$. Similarly, $L_{i}(\mu)$ is continuous in $\mu$, and $L_{i}(0)=0$. Therefore, we have
\begin{eqnarray*}
\left| B_i(\bm{\theta}^{t+1})-B_i(\bm{\theta}^t) \right| &\leq& \textstyle \sum_{j=1}^N \left|\theta_j^{t+1}-\theta_j^t\right| \cdot L_{ij}(\mu) \\
&\leq& L_i(\mu) \cdot \left\| \bm{\theta}^{t+1}-\bm{\theta}^t \right\|_1.
\end{eqnarray*}
\end{IEEEproof}

\section{Proof of Theorem~\ref{theorem:Convergence}}\label{proof:Convergence}

Before we prove the convergence of the updates \eqref{eqn:UpdateEffort}--\eqref{eqn:UpdateBelief}, we show that \emph{if} they converge to a triple $\{\theta_i^*,e_i^*,\beta_i^*\}_{i \in \mathcal{N}}$, this triple $\{\theta_i^*,e_i^*,\beta_i^*\}_{i \in \mathcal{N}}$ is a CE. Note that when the updates converge, each agent $i$ has a fixed rating $\theta_i^{t+1} = \theta_i^t = \theta_i^*$, a fixed effort level $e_i^t = e_i^*$, and a fixed payoff $u_i\left(m,\theta_i^t, d(\bm{\theta}^t), \bm{e}^t\right) = u_i\left(m,\theta_i^*, d(\bm{\theta}^*), \bm{e}^*\right)$. First, the best response \eqref{eqn:UpdateEffort} becomes
\begin{eqnarray*}
e_i^* = \arg\max_{e_i\in [0, e_i^{\rm max}]} & &  \!\!\!\!\!\!\!\!\!\!\! (1-\delta_i) \cdot u_i(m, \theta_i^*, d(\bm{\theta}^*), e_i, \bm{e}_{-i}^*) \\
&+& \delta_i \cdot f_i(\alpha_i,\beta_i^*, \theta_i^*, d^*, e_i),
\end{eqnarray*}
which is exactly the first requirement of ``incentive compatibility'' in the definition of CE. Second, the rating update \eqref{eqn:UpdateRating} ensures $\theta_i^* = q_i(e_i^*)$, which fulfills the second requirement of ``stable and correct rating'' in the definition of CE. Finally, the update of $\beta$ in \eqref{eqn:UpdateBelief} becomes
\begin{eqnarray*}
\beta_i^*  = u_i\left(m,\theta_i^*, d(\bm{\theta}^*), \bm{e}^*\right) - \alpha_i \cdot \bar{b}(\theta_i, d(\bm{\theta}^*), e_i^*).
\end{eqnarray*}
Rearranging terms and using the definition of the conjecture function \eqref{eqn:conjecture}, we have
$$
f_i(\alpha_i,\beta_i^*, \theta_i^*, d(\bm{\theta}^*), e_i^*) = u_i\left(m,\theta_i^*, d(\bm{\theta}^*), \bm{e}^*\right). 
$$
In other words, the conjecture is indeed equal to the true payoff, which fulfills the third requirement of ``correct conjectures'' in the definition of CE.

Next, we prove that the updates \eqref{eqn:UpdateEffort}--\eqref{eqn:UpdateBelief} \emph{do} converge under a small enough step size $\mu$ in the rating update rule. Consider a sequence of rating profiles $\left\{\bm{\theta}^t\right\}_{t=0}^\infty$ generated by the updates \eqref{eqn:UpdateEffort}--\eqref{eqn:UpdateBelief}. Our goal is to prove that there exists a $\rho \in (0,1)$ such that 
$$
\left\|\bm{\theta}^{t+2}-\bm{\theta}^{t+1}\right\|_1 \leq \rho \cdot \left\|\bm{\theta}^{t+1}-\bm{\theta}^{t}\right\|_1
$$ 
for all $t \geq 0$. If the above is true, the sequence $\left\{\bm{\theta}^t\right\}_{t=0}^\infty$ will be a Cauchy sequence in $\mathbb{R}_+^N$, and hence will converge.

We prove the contraction property of the differences between consecutive rating profiles.
Note that the rating update rule \eqref{eqn:UpdateRating} is asynchronous. Hence, agent $i$'s rating will not be updated if it did not grade an assignment. Recall Lemma \ref{Lemma: Properties of best response}. For each agent $i$, there are two cases:
\begin{itemize}
\item If there is rating update at time $t+1$, we have 
$$
\left| \bm{\theta}^{t+2}-\bm{\theta}^{t+1} \right| \leq L_i(\mu) \cdot \left\| \bm{\theta}^{t+1}-\bm{\theta}^t \right\|_1,
$$
where $L_{i}(\mu)$ is as in Lemma \ref{Lemma: Properties of best response}.
\item If there is no rating update at time $t+1$, we have $\theta_i^{t+2}=\theta_i^{t+1}$, and hence
$$
\left| \bm{\theta}^{t+2}-\bm{\theta}^{t+1} \right| = 0 \leq L_i(\mu) \cdot \left\| \bm{\theta}^{t+1}-\bm{\theta}^t \right\|_1.
$$
\end{itemize}
Hence, the above inequality holds despite the asynchronous rating update.

Moreover, since $q_i$ is concave and thus has a decreasing derivative, we have 
\begin{eqnarray*}
\left| q_i\left( B_i(\bm{\theta}^{t+1}) \right) - q_i\left( B_i(\bm{\theta}^t) \right) \right|
&\leq& q_i^\prime(0) \left| B_i(\bm{\theta}^{t+1})-B_i(\bm{\theta}^t) \right| \\
&\leq& q_i^\prime(0) L_i(\mu) \cdot \left\| \bm{\theta}^{t+1}-\bm{\theta}^t \right\|_1.
\end{eqnarray*}

As a result, we have
\begin{eqnarray*}
& & \left\|\bm{\theta}^{t+2} - \bm{\theta}^{t+1}\right\|_1 = \sum_{i=1}^N \left|\theta_i^{t+2} - \theta_i^{t+1}\right| \\
& \leq & \sum_{i=1}^N \left[ (1-\mu) \cdot \left|\theta_i^t - \theta_i^{t-1}\right| + \mu \left| q_i\left( B_i(\bm{\theta}^{t+1}) \right) - q_i\left( B_i(\bm{\theta}^t) \right) \right| \right] \\
& \leq & (1-\mu) \left\|\bm{\theta}^t - \bm{\theta}^{t-1}\right\|_1 + \mu \cdot \left(\sum_{i=1}^N q_i^\prime(0) L_i(\mu)\right) \cdot \left\|\bm{\theta}^t - \bm{\theta}^{t-1}\right\|_1 \\
& = & \left[ (1-\mu) + \mu \cdot \left(\sum_{i=1}^N q_i^\prime(0) L_i(\mu)\right) \right] \cdot \left\|\bm{\theta}^t - \bm{\theta}^{t-1}\right\|_1.
\end{eqnarray*}
Due to property 3 of Lemma~\ref{Lemma: Properties of best response}, we
can find a small enough step size $\mu>0$ such that $\sum_{i=1}^N q_i^\prime(0) L_i(\mu)<1$, and hence $\rho \triangleq (1-\mu) + \mu \cdot \left(\sum_{i=1}^N q_i^\prime(0) L_i(\mu)\right) <1$.

\section{Proof of Results in Section~\ref{sec:design}}\label{proof:section_design}

\emph{Proof of Proposition 1:} We analyze the best response defined in \eqref{eqn:best_response}. According to \eqref{eqn:best_response}, the best response maximizes an objective function that consists of two terms. Under a matching rule that is independent of the author's rating, the second term of the objective function in \eqref{eqn:best_response}, namely 
$$
\textstyle\delta_i \alpha_i   \sum_{k_j \neq k_i^+}  m_{k_i^+ k_j}\left( (1-\mu) \theta_i^t + \mu q_i(e_i), \theta_j^t \right) \cdot b_i(\theta_j^t),
$$
is independent of $i$'s rating, and hence is independent of its effort level. Hence, $i$'s best response is an effort level that maximizes the first term of the objective function, namely 
$$
- (1-\delta_i) M c_i(e_i).
$$ 
Since each agent $i$'s cost is strictly increasing in its effort level, its best response is then $e_i^t=0$ under any rating profile $\bm{\theta}^t$. Since student $i$ chooses $e_i^t=0$ at any time $t$, its rating will converge to $\theta_i^*=q_i(0)=0$. \hfill $\Box$

\emph{Proof of Proposition~\ref{proposition:initial_rating}:} Given the same { current} initial rating $\theta^0$ of everyone, { based on Property 1) of the matching rule, agent $i$ is matched to a reviewer with the same rating $\theta^0$ with probability $1$. Hence, agent $i$ incurs a current cost of $-c_i(e_i)$ by choosing effort $e_i$. 

Next, we look at agent $i$'s conjectured future payoff, which depends on the new rating $(1-\mu) \theta^0 + \mu q_i(e_i)$ when agent $i$ exerts effort $e_i$. If $(1-\mu) \theta^0 + \mu q_i(e_i) \geq \theta^0$, namely $q_i(e_i) \geq \theta^0$, based on Properties 2)-a) and 2)-c) of the matching rule, agent $i$ is matched to a reviewer with rating $\theta^0$ with probability $1$. If $(1-\mu) \theta^0 + \mu q_i(e_i) < \theta^0$, namely $q_i(e_i) < \theta^0$, based on Property 2)-b) of the matching rule, $i$ is matched to a reviewer with rating $\theta^0$ with probability $\frac{(1-\mu) \cdot \theta^0+\mu \cdot q_i(e_i)}{\theta^0}$. In summary,} $i$'s belief function at time $0$, $f_i(\alpha_i,\beta_i,\theta^0,d(\bm{\theta}^0),e_i)$, can be calculated as
\begin{eqnarray*}
\left\{\begin{array}{ll} \alpha_i \cdot b_i(\theta^0) + \beta_i, & \mathrm{when}~q_i(e_i) \geq \theta^0 \\ \alpha_i \cdot \frac{(1-\mu) \cdot \theta^0+\mu \cdot q_i(e_i)}{\theta^0} \cdot b_i(\theta^0) + \beta_i, & \mathrm{when}~q_i(e_i) < \theta^0 \end{array}\right..
\end{eqnarray*}

At time $0$, agent $i$ chooses its effort level $e_i^0$ by solving the optimization problem \eqref{eqn:best_response}, which is equivalent to solving
$$
e_i^0 = \arg\max_{e_i} \left\{ -(1-\delta_i) c_i(e_i) + \delta_i f_i(\alpha_i,\beta_i,\theta^0,d(\bm{\theta}^0),e_i) \right\}.
$$

Since $f_i(\alpha_i,\beta_i,\theta^0,d(\bm{\theta}^0),e_i)$ is a constant and does not increase with the effort level $e_i$ when $q_i(e_i) \geq \theta^0$, we restrict ourselves (without loss of generality) to $q_i(e_i) \leq \theta^0$. { To find the solution $e_i^0$, we look at the derivative of the objective with respect to $e_i$ for any $\theta^0>0$ and $e_i \in [0, q_i^{-1}(\theta^0))$:
$$
F_i(e_i, \theta^0) \triangleq -(1-\delta_i) c_i^\prime(e_i) + \delta_i \alpha_i \mu q_i^\prime( e_i ) \frac{b_i(\theta^0)}{\theta^0}.
$$
The left derivative of the objective with respect to $e_i$ when $e_i=q_i^{-1}(\theta^0)$ is 
\begin{eqnarray*}
& & F_i(q_i^{-1}(\theta^0), \theta^0) 
\triangleq \lim_{e_i \uparrow q_i^{-1}(\theta^0)} F_i(e_i, \theta^0) \\
&=& -(1-\delta_i) c_i^\prime(q_i^{-1}(\theta^0)) + \delta_i \alpha_i \mu q_i^\prime( q_i^{-1}(\theta^0) ) \frac{b_i(\theta^0)}{\theta^0}.
\end{eqnarray*}

Since $q_i(e_i)$ is strictly increasing in $e_i$, $q_i^{-1}(\theta^0)$ is strictly increasing in $\theta^0$. Due to strict convexity of $c_i$, $c_i^\prime(q_i^{-1}(\theta^0))$ is strictly increasing in $\theta^0$. Since $q_i$ is concave and strictly increasing in $e_i$, $q_i^\prime( q_i^{-1}(\theta^0) )$ is positive and decreasing in $\theta^0$. Since $b_i$ is concave and strictly increasing in $\theta^0$, $\frac{b_i(\theta^0)}{\theta^0}$ is positive and decreasing in $\theta^0$. As a result, $F_i(q_i^{-1}(\theta^0), \theta^0) $ is strictly decreasing in $\theta^0$.

Moreover, due to continuous differentiability assumptions of $c_i$, $q_i$, and $b_i$, $F_i(q_i^{-1}(\theta^0), \theta^0) $ is continuous in $\theta^0$.

In addition, due to our assumptions, $c_i^\prime(0)=0$, $q_i^\prime(0)>0$, and $\lim_{\theta^0 \downarrow 0} \frac{b_i(\theta^0)}{\theta^0} = b_i^\prime(0) > 0$. Hence, we have $F_i(q_i^{-1}(0), 0) > 0$.

Since $F_i(q_i^{-1}(\theta^0), \theta^0) $ is strictly decreasing and continuous in $\theta^0$, and since $F_i(q_i^{-1}(0), 0) > 0$, there exists a small enough $\underline{\theta}_i$ such that $F_i(q_i^{-1}(\theta^0), \theta^0) \geq 0$ for any $\theta^0 \leq \underline{\theta}_i$. We define $\underline{\theta} \triangleq \max_{i} \underline{\theta}_i$. Then for any $\theta^0 \leq \underline{\theta}$, we have $F_i(q_i^{-1}(\theta^0), \theta^0) \geq 0$ for all $i$. 

For any $\theta^0 \leq \underline{\theta}$, due to strict concavity of the objective function, the derivative of the objective with respect to $e_i$, $F_i(e_i, \theta^0)$, is decreasing in $e_i$ and thus satisfies $F_i(e_i, \theta^0) \geq 0$ for all $e_i \leq q_i^{-1}(\theta^0)$. Therefore, each agent $i$ chooses } effort $e_i^0 = q_i^{-1}(\theta^0)$ at time $0$, and gets the same rating $\theta_i^1=\theta^0$ at time $1$, under which it chooses the same effort $e_i^1 = q_i^{-1}(\theta^0)$. As a result, the system will stay the same. Hence, the equilibrium rating is $\theta^0$, and the equilibrium effort level is $e_i^* = q_i^{-1}(\theta^0)$. \hfill $\Box$

\emph{Proof of Theorem~\ref{theorem:convergence_baseline}:}

\emph{Claims 1-2:} First, we prove the first claim that the baseline matching rule is a desirable rule. Under the large population assumption (i.e., Assumption~\ref{assumption:Population}), there are multiple agents of the same type. These agents of the same type will choose the same effort level and hence have the same rating. According to Property 1) in the baseline matching rule, they will always have exactly one product (from another student of the same rating) to review, namely $M=1$.

It remains to show that the conjectured expected benefit is increasing and concave in the effort level. Under the baseline matching rule, the conjectured expected benefit can be explicitly computed as follows:
\begin{eqnarray}\label{eqn:ExpectedBenefit_Baseline}
&     & \textstyle \sum_{k_j \neq k_i} \left[ m_{k_i k_j}(\theta_i,d(\bm{\theta})_{k_j}) \cdot b_i(d(\bm{\theta})_{k_j}) \right] \\
& = & \frac{\theta_i-d(\bm{\theta})_{k_i+1}}{d(\bm{\theta})_{k_i-1}-d(\bm{\theta})_{k_i+1}} \cdot b_i(d(\bm{\theta})_{k_i-1}) \nonumber \\
& + & \frac{d(\bm{\theta})_{k_i-1}-\theta_i}{d(\bm{\theta})_{k_i-1}-d(\bm{\theta})_{k_i+1}} \cdot b_i(d(\bm{\theta})_{k_i+1}) \nonumber \\
& = & \frac{b_i(d(\bm{\theta})_{k_i-1})-b_i(d(\bm{\theta})_{k_i+1})}{d(\bm{\theta})_{k_i-1}-d(\bm{\theta})_{k_i+1}} \cdot \theta_i \nonumber \\
& + & \frac{d(\bm{\theta})_{k_i-1} \cdot b_i(d(\bm{\theta})_{k_i+1})-d(\bm{\theta})_{k_i+1} \cdot b_i(d(\bm{\theta})_{k_i-1})}{d(\bm{\theta})_{k_i-1}-d(\bm{\theta})_{k_i+1}}. \nonumber
\end{eqnarray}
We can see that the expected benefit is a piecewise linear function of $\theta_i$. When $i$ is ranked at $k_i$, the slope of the function is $\frac{b_i(d(\bm{\theta})_{k_i-1})-b_i(d(\bm{\theta})_{k_i+1})}{d(\bm{\theta})_{k_i-1}-d(\bm{\theta})_{k_i+1}}$. Since $b_i$ is an increasing function of $\theta$, we have
$
\frac{b_i(d(\bm{\theta})_{k_i-1})-b_i(d(\bm{\theta})_{k_i+1})}{d(\bm{\theta})_{k_i-1}-d(\bm{\theta})_{k_i+1}} \geq 0
$
for all $k_i$. Hence, the expected benefit is increasing in $\theta_i$.

Since $b_i$ is a concave and increasing function of $\theta$, we have
$
\frac{b_i(d(\bm{\theta})_{k_i-1})-b_i(d(\bm{\theta})_{k_i+1})}{d(\bm{\theta})_{k_i-1}-d(\bm{\theta})_{k_i+1}} \leq \frac{b_i(d(\bm{\theta})_{k_i})-b_i(d(\bm{\theta})_{k_i+2})}{d(\bm{\theta})_{k_i}-d(\bm{\theta})_{k_i+2}}.
$
Hence, the slope decreases when $i$ is ranked higher. Therefore, the expected benefit is concave in $\theta_i$. Since $(1-\mu) \theta_i + \mu q_i(e_i)$ is increasing and concave in $e_i$, the expected benefit
$$
\textstyle \sum_{k_j \neq k_i} \left[ m_{k_i j}((1-\mu) \theta_i + \mu q_i(e_i), d(\bm{\theta})_{k_j}) \cdot b_i(d(\bm{\theta})_{k_j}) \right]
$$
is increasing and concave in the effort level $e_i$.

In summary, the baseline matching rule is a desirable rule. The second claim on the convergence then follows due to Theorem 1.

\emph{Claim 3:} Now we prove the third claim that a more capable agent always gets no lower ratings than a less capable agent. More specifically, we will prove that as long as a more capable agent has a no lower current rating, he/she will exert effort high enough such that his/her next rating is no lower than that of a less capable agent. Then under the same initial rating for all the agents, the third claim follows. Our proof strategy is to first derive the sufficient and necessary conditions for the best response effort levels, and then show (by contradiction) that if a less capable agent exerts a high effort level that results in a next rating higher than that of a more capable agent, this high effort level violates the conditions and cannot be the best response. The main technical difficulty arises since the best responses may be at the smooth or non-smooth points of the objective function, which results in different sufficient and necessary conditions and needs separate treatments.

\emph{Sufficient and necessary conditions for best responses:} First, we look at $i$'s best response at time $t$:
\begin{multline*}
e_i^t = \arg \max_{e_i \in [0, e_i^{\rm max}]}
- (1-\delta_i) c_i(e_i) + \\
\delta_i \alpha_i   \sum_{k_j \neq k_i}  m_{k_i k_j}\left( (1-\mu) \theta_i^t + \mu q_i(e_i), d(\bm{\theta}^t)_{k_j} \right) \cdot b_i\left(d(\bm{\theta}^t)_{k_j} \right).
\end{multline*}
Under the baseline rule, if the next rating $\theta_i^{t+1}=(1-\mu) \theta_i^t + \mu q_i(e_i)$ lies between $d(\bm{\theta}^t)_k$ and $d(\bm{\theta}^t)_{k+1}$ for some $k\in\{1,\ldots,N-1\}$, the expected benefit is calculated as:
\begin{eqnarray*}
&     & \sum_{k_j \neq k_i}  m_{k_i k_j}\left( (1-\mu) \theta_i^t + \mu q_i(e_i), d(\bm{\theta}^t)_{k_j} \right) \cdot b_i\left(d(\bm{\theta}^t)_{k_j} \right) \\
& = & \frac{b_i(d(\bm{\theta}^t)_{k})-b_i(d(\bm{\theta})_{k+1}^t)}{d(\bm{\theta}^t)_{k}-d(\bm{\theta}^t)_{k+1}} \cdot \left[ (1-\mu) \theta_i^t + \mu q_i(e_i) \right] \nonumber \\
& + & \frac{d(\bm{\theta}^t)_{k} \cdot b_i(d(\bm{\theta}^t)_{k+1})-d(\bm{\theta}^t)_{k+1} \cdot b_i(d(\bm{\theta}^t)_{k})}{d(\bm{\theta}^t)_{k}-d(\bm{\theta}^t)_{k+1}}. \nonumber
\end{eqnarray*} 
Removing all the terms that are unrelated to $e_i$, the best response can be simplified into:
\begin{eqnarray}\label{eqn:best_response_baseline}
e_i^t &=& \arg \max_{e_i \in [0, e_i^{\rm max}]}
- (1-\delta_i) c_i(e_i) \\
& & ~~~~~~~~ + \delta_i \alpha_i \frac{b_i(d(\bm{\theta}^t)_{k})-b_i(d(\bm{\theta})_{k+1}^t)}{d(\bm{\theta}^t)_{k}-d(\bm{\theta}^t)_{k+1}} \cdot \mu q_i(e_i). \nonumber
\end{eqnarray}

Recall that the expected benefit is piecewise linear in $\theta_i^t$. Hence, the objective function in \eqref{eqn:best_response_baseline} is a piecewise smooth function of $e_i$. There are $N$ nonsmooth points where the effort $e_i$ satisfies $\theta_i^{t+1}=(1-\mu)\theta_i^t+\mu q_i(e_i)=d(\bm{\theta}^t)_k$ for some $k \in \{1,\ldots,N\}$.

Since the objective function is strictly concave, the best response $e_i^t$ should satisfy the following conditions. When $e_i^t$ results in a nonsmooth point, namely $\theta_i^{t+1}=(1-\mu)\theta_i^t+\mu q_i(e_i^t)=d(\bm{\theta}^t)_k$, the left derivative at $e_i^t$ must be nonnegative and the right derivative at $e_i^t$ must be nonpositive, i.e.:
\begin{eqnarray}\label{eqn:FOC_nonsmooth_left}
(1-\delta_i) c_i^\prime(e_i^t) \leq
\delta_i \alpha_i \mu q_i^\prime(e_i^t) \cdot \frac{b_i\left( d(\bm{\theta}^t)_k \right)-b_i\left( d(\bm{\theta}^t)_{k+1} \right)}{d(\bm{\theta}^t)_k-d(\bm{\theta}^t)_{k+1}}, \nonumber
\end{eqnarray}
and
\begin{eqnarray}\label{eqn:FOC_nonsmooth_right}
(1-\delta_i) c_i^\prime(e_i^t) \geq
+ \delta_i \alpha_i \mu q_i^\prime(e_i^t) \cdot \frac{b_i\left( d(\bm{\theta}^t)_{k-1} \right)-b_i\left( d(\bm{\theta}^t)_{k} \right)}{d(\bm{\theta}^t)_{k-1}-d(\bm{\theta}^t)_{k}}. \nonumber 
\end{eqnarray}

When $e_i^t$ results in a smooth point, namely $(1-\mu)\theta_i^t+\mu q_i(e_i^t) \in \left( d(\bm{\theta}^t)_k, d(\bm{\theta}^t)_{k+1} \right)$, the derivative of the objective in \eqref{eqn:best_response_baseline} exists and should be $0$, namely
\begin{eqnarray}\label{eqn:FOC_smooth}
0 &=& -(1-\delta_i) c_i^\prime(e_i^t) \\
& & + \delta_i \alpha_i \mu q_i^\prime(e_i^t) \cdot \frac{b_i\left( d(\bm{\theta}^t)_k \right)-b_i\left( d(\bm{\theta}^t)_{k+1} \right)}{d(\bm{\theta}^t)_k-d(\bm{\theta}^t)_{k+1}}. \nonumber 
\end{eqnarray}

\emph{Proof by contradiction:}
Now pick two arbitrary agents $i$ and $j$ with $i$ being more capable than $j$. We prove that if $\theta_i^t \geq \theta_j^t$, their best responses must satisfy 
$$
(1-\mu) \theta_i^t + \mu q_i(e_i^t) \geq (1-\mu) \theta_j^t + \mu q_j(e_j^t).
$$
We distinguish two cases depending on whether $e_i^t$ results in a smooth point or not. In each case, we further separate our discussions into two subcases depending on whether $e_j^t$ results in a smooth point or not.
\begin{itemize}
\item Case 1: $e_i^t$ results in a smooth point, namely $(1-\mu)\theta_i^t+\mu q_i(e_i^t) \in \left( d(\bm{\theta}^t)_k, d(\bm{\theta}^t)_{k+1} \right)$. Then $e_i^t$ must satisfy \eqref{eqn:FOC_smooth}, which is equivalent to
$$
\frac{ (1-\delta_i) c_i^\prime(e_i^t) }{ \delta_i \alpha_i \mu q_i^\prime(e_i^t) } = \frac{b_i\left( d(\bm{\theta}^t)_k \right)-b_i\left( d(\bm{\theta}^t)_{k+1} \right)}{d(\bm{\theta}^t)_k-d(\bm{\theta}^t)_{k+1}}.
$$
Suppose that $j$ chooses a best response $e_j^t$ such that
$$
(1-\mu) \theta_i^t + \mu q_i(e_i^t) < (1-\mu) \theta_j^t + \mu q_j(e_j^t).
$$
Since $\theta_j^t \leq \theta_i^t$, and since $q_j(e) < q_i(e)$, we must have $e_j^t > e_i^t$, and hence $\frac{ (1-\delta_j) c_j^\prime(e_j^t) }{ \delta_j \alpha_j q_j^\prime(e_j^t) } > \frac{ (1-\delta_j) c_j^\prime(e_i^t) }{ \delta_j \alpha_j q_j^\prime(e_i^t) } \geq \frac{ (1-\delta_i) c_i^\prime(e_i^t) }{ \delta_i \alpha_i q_i^\prime(e_i^t) }$. Moreover, for any $k^\prime \leq k$, we have
\begin{eqnarray*}
&\frac{b_j\left( d(\bm{\theta}^t)_{k^\prime} \right)-b_j\left( d(\bm{\theta}^t)_{k^\prime+1} \right)}{d(\bm{\theta}^t)_{k^\prime}-d(\bm{\theta}^t)_{k^\prime+1}} \leq \frac{b_i\left( d(\bm{\theta}^t)_{k} \right)-b_i\left( d(\bm{\theta}^t)_{k+1} \right)}{d(\bm{\theta}^t)_{k}-d(\bm{\theta}^t)_{k+1}},
\end{eqnarray*}
which leads to
\begin{eqnarray}\label{eqn:contradiction1}
\frac{ (1-\delta_j) c_j^\prime(e_j^t) }{ \delta_j \alpha_j \mu q_j^\prime(e_j^t) } > \frac{b_j\left( d(\bm{\theta}^t)_{k^\prime} \right)-b_j\left( d(\bm{\theta}^t)_{k^\prime+1} \right)}{d(\bm{\theta}^t)_{k^\prime}-d(\bm{\theta}^t)_{k^\prime+1}},
\end{eqnarray}
for any $k^\prime \leq k$.

\begin{itemize}
\item Subcase 1:
If $e_j^t$ results in any smooth point $(1-\mu)\theta_j^t+\mu q_j(e_j^t) \in \left( d(\bm{\theta}^t)_{k^\prime}, d(\bm{\theta}^t)_{k^\prime+1} \right)$ with $k^\prime \leq k$, then \eqref{eqn:contradiction1} violates the condition \eqref{eqn:FOC_smooth}.

\item Subcase 2:
If $e_j^t$ results in any nonsmooth point $(1-\mu)\theta_j^t+\mu q_j(e_j^t) = d(\bm{\theta}^t)_{k^\prime}$ with $k^\prime \leq k$, then \eqref{eqn:contradiction1} violates the optimality condition \eqref{eqn:FOC_nonsmooth_left}.
\end{itemize}

In summary, in both subcases, $e_j^t$ cannot be the best response, which leads to contradiction. Hence, in Case 1 we must have 
$$
(1-\mu) \theta_i^t + \mu q_i(e_i^t) > (1-\mu) \theta_j^t + \mu q_j(e_j^t).
$$

\item Case 2: $e_i^t$ results in a nonsmooth point, namely $(1-\mu)\theta_i^t+\mu q_i(e_i^t)=d(\bm{\theta}^t)_k$. Then $e_i^t$ must satisfy \eqref{eqn:FOC_nonsmooth_right}, which leads to
$$
\frac{ (1-\delta_i) c_i^\prime(e_i^t) }{ \delta_i \alpha_i \mu q_i^\prime(e_i^t) } \geq \frac{b_i\left( d(\bm{\theta}^t)_{k-1} \right)-b_i\left( d(\bm{\theta}^t)_{k} \right)}{d(\bm{\theta}^t)_{k-1}-d(\bm{\theta}^t)_{k}}.
$$
Suppose that $j$ chooses a best response $e_j^t$ such that
$$
(1-\mu) \theta_i^t + \mu q_i(e_i^t) < (1-\mu) \theta_j^t + \mu q_j(e_j^t).
$$
Then we must have $e_j^t > e_i^t$. Following the same logic that leads to \eqref{eqn:contradiction1}, we have
\begin{eqnarray}\label{eqn:contradiction2}
\frac{ (1-\delta_j) c_j^\prime(e_j^t) }{ \delta_j \alpha_j \mu q_j^\prime(e_j^t) } > \frac{b_j\left( d(\bm{\theta}^t)_{k^\prime} \right)-b_j\left( d(\bm{\theta}^t)_{k^\prime+1} \right)}{d(\bm{\theta}^t)_{k^\prime}-d(\bm{\theta}^t)_{k^\prime+1}},
\end{eqnarray}
for any $k^\prime \leq k-1$.

Similar to Case 1, since $(1-\mu) \theta_j^t + \mu q_j(e_i^t)e_j^t > d(\bm{\theta}^t)_{k}$, we have two subcases: $(1-\mu) \theta_j^t + \mu q_j(e_i^t)e_j^t = d(\bm{\theta}^t)_{k^\prime}$, and $(1-\mu) \theta_j^t + \mu q_j(e_i^t)e_j^t \in \left( d(\bm{\theta}^t)_{k^\prime}, d(\bm{\theta}^t)_{k^\prime+1} \right)$, where $k^\prime \leq k-1$. Since the inequality \eqref{eqn:contradiction2} violates \eqref{eqn:FOC_nonsmooth_left} and \eqref{eqn:FOC_smooth}, $e_j^t$ cannot be the best response in either subcase.

Hence, in Case 2 we must have 
$$
(1-\mu) \theta_i^t + \mu q_i(e_i^t) \geq (1-\mu) \theta_j^t + \mu q_j(e_j^t).
$$
\end{itemize}

We have shown that if a more capable agent has a current rating no lower than that of a less capable agent, then its next rating is no lower. Under the same initial rating, more capable agents always have no lower ratings.
\hfill $\Box$

\emph{Proof of Theorem~\ref{theorem:extensions}:}
A rating profile $\bm{\theta}^*$ is an equilibrium rating profile if each student $i$'s effort $e_i^* = q_i^{-1}(\theta_i^*)$ is the best response. Under the baseline matching rule, we have derived the sufficient and necessary conditions for $e_i^*$ to be the best response in \eqref{eqn:FOC_nonsmooth_left}--\eqref{eqn:FOC_smooth}. Hence, a rating profile $\bm{\theta}^*$ {and the associated effort profile $\bm{e}^*$ with $e_i^*=q_i^{-1}(\theta_i^*)$} are an equilibrium rating profile and the associated equilibrium effort profile if and only if: when $k_i=1$,
\begin{eqnarray}\label{eqn:baseline_1}
(1-\delta_i) c_i^\prime(e_i^*) \leq \textstyle\delta_i \alpha_i \mu q_i^\prime(e_i^*) \cdot \frac{b_i(d(\bm{\theta}^*)_{k_i})-b_i(d(\bm{\theta}^*)_{k_i+1})}{d(\bm{\theta}^*)_{k_i}-d(\bm{\theta}^*)_{k_i+1}}; 
\end{eqnarray}
when $k_i\in\{2,\ldots,N-1\}$, 
\begin{eqnarray}\label{eqn:baseline_k_left}
(1-\delta_i) c_i^\prime(e_i^*) \leq
\textstyle \delta_i \alpha_i \mu q_i^\prime(e_i^*) \cdot \frac{b_i(d(\bm{\theta}^*)_{k_i})-b_i(d(\bm{\theta}^*)_{k_i+1})}{d(\bm{\theta}^*)_{k_i}-d(\bm{\theta}^*)_{k_i+1}}, 
\end{eqnarray}
\begin{eqnarray}\label{eqn:baseline_k_right}
(1-\delta_i) c_i^\prime(e_i^*) \geq
\textstyle \delta_i \alpha_i \mu q_i^\prime(e_i^*) \cdot \frac{b_i(d(\bm{\theta}^*)_{k_i-1})-b_i(d(\bm{\theta}^*)_{k_i})}{d(\bm{\theta}^*)_{k_i-1}-d(\bm{\theta}^*)_{k_i}}; 
\end{eqnarray}
and when $k_i=N$,
\begin{eqnarray}\label{eqn:baseline_N}
(1-\delta_i) c_i^\prime(e_i^*) \geq
\textstyle \delta_i \alpha_i \mu q_i^\prime(e_i^*) \cdot \frac{b_i(d(\bm{\theta}^*)_{k_i-1})-b_i(d(\bm{\theta}^*)_{k_i})}{d(\bm{\theta}^*)_{k_i-1}-d(\bm{\theta}^*)_{k_i}}. 
\end{eqnarray}

Following the same procedure, we can show that under the asymmetric extension with $\gamma>0$, a rating profile $\hat{\bm{\theta}}^*$ {and the associated effort profile $\hat{\bm{e}}^*$ with $\hat{e}_i^*=q_i^{-1}(\hat{\theta}_i^*)$} are an equilibrium rating profile and the associated equilibrium effort profile if and only if: when $k_i=1$,
\begin{eqnarray}\label{eqn:first_1}
(1-\delta_i) c_i^\prime(\hat{e}_i^*) \leq
\textstyle \delta_i \alpha_i \mu q_i^\prime(\hat{e}_i^*) \left[ \frac{b_i(d(\bm{\hat{\theta}}^*)_{k_i})-b_i(d(\bm{\hat{\theta}}^*)_{k_i+1})}{d(\bm{\hat{\theta}}^*)_{k_i}-d(\bm{\hat{\theta}}^*)_{k_i+1}} + \gamma \right]\!\!
\end{eqnarray}
when $k_i\in\{2,\ldots,N-1\}$,
\begin{eqnarray}\label{eqn:first_k_left}
(1-\delta_i) c_i^\prime(\hat{e}_i^*) \leq
\textstyle \delta_i \alpha_i \mu q_i^\prime(\hat{e}_i^*) \left[ \frac{b_i(d(\bm{\hat{\theta}}^*)_{k_i})-b_i(d(\bm{\hat{\theta}}^*)_{k_i+1})}{d(\bm{\hat{\theta}}^*)_{k_i}-d(\bm{\hat{\theta}}^*)_{k_i+1}} + \gamma \right]\!\!
\end{eqnarray}
\begin{eqnarray}\label{eqn:first_k_right}
(1-\delta_i) c_i^\prime(\hat{e}_i^*) \geq
\textstyle \delta_i \alpha_i \mu q_i^\prime(\hat{e}_i^*) \left[ \frac{b_i(d(\bm{\hat{\theta}}^*)_{k_i-1})-b_i(d(\bm{\hat{\theta}}^*)_{k_i})}{d(\bm{\hat{\theta}}^*)_{k_i-1}-d(\bm{\hat{\theta}}^*)_{k_i}} + \gamma \right]\!\!
\end{eqnarray}
and when $k_i=N$,
\begin{eqnarray}\label{eqn:first_N}
(1-\delta_i) c_i^\prime(\hat{e}_i^*) \geq
\textstyle \delta_i \alpha_i \mu q_i^\prime(\hat{e}_i^*) \left[ \frac{b_i(d(\bm{\hat{\theta}}^*)_{k_i-1})-b_i(d(\bm{\hat{\theta}}^*)_{k_i})}{d(\bm{\hat{\theta}}^*)_{k_i-1}-d(\bm{\hat{\theta}}^*)_{k_i}} + \gamma \right]\!\!
\end{eqnarray}

Observe that the defining inequalities for the baseline matching rule and those for the first extension are identical up to the additive term $\gamma$.
For any equilibrium rating profile $\bm{\theta}^*$ under the baseline rule $m$, we can define $\hat{\bm{\theta}}^* = \bm{\theta}^* + \varepsilon \cdot \bm{1}_N$, where $\varepsilon>0$ is a small constant. In the following we show that a strictly positive and small enough $\varepsilon>0$ leads to a strictly better equilibrium rating profile $\hat{\bm{\theta}}^*>\bm{\theta}^*$. 

Since $\hat{\theta}_i^* = q_i(\hat{e}_i^*)$ and $\theta_i^* = q_i(e_i^*)$, and since $q_i$ is strictly increasing in $e_i$, we have $\hat{e}_i^* = e_i^* + \varepsilon_i$, where $\varepsilon_i>0$. Since $c_i$ is convex and $q_i$ is concave, we have $c_i^\prime(\hat{e}_i^*)>c_i^\prime(e_i^*)$ and $q_i^\prime(\hat{e}_i^*)<q_i^\prime(e_i^*)$. Since $b_i$ is concave and increasing, we have 
\begin{eqnarray*}
\frac{b_i(d(\bm{\hat{\theta}}^*)_{k_i-1})-b_i(d(\bm{\hat{\theta}}^*)_{k_i})}{d(\bm{\hat{\theta}}^*)_{k_i-1}-d(\bm{\hat{\theta}}^*)_{k_i}} < \frac{b_i(d(\bm{\theta}^*)_{k_i-1})-b_i(d(\bm{\theta}^*)_{k_i})}{d(\bm{\theta}^*)_{k_i-1}-d(\bm{\theta}^*)_{k_i}}
\end{eqnarray*}
for any $k_i=2,\ldots,N$. Therefore, {if the inequalities with ``$\geq$'', i.e., \eqref{eqn:baseline_k_right} and \eqref{eqn:baseline_N}, hold for $\bm{\theta}^*$ and $\bm{e}^*$, then the inequalities with ``$\geq$'', i.e., \eqref{eqn:first_k_right} and \eqref{eqn:first_N}, hold for $\hat{\bm{\theta}}^*$ and $\hat{\bm{e}}^*$ as long as $\gamma>0$ is small enough. For any such $\gamma$, if the inequalities with ``$\leq$'', i.e., \eqref{eqn:baseline_1} and \eqref{eqn:baseline_k_left}, hold for $\bm{\theta}^*$ and $\bm{e}^*$, then the inequalities with ``$\leq$'', i.e., \eqref{eqn:first_1} and \eqref{eqn:first_k_left} hold for $\hat{\bm{\theta}}^*$ and $\hat{\bm{e}}^*$ as long as $\varepsilon_i(\gamma)>0$ is small enough.} In conclusion, we can find an equilibrium rating profile $\hat{\bm{\theta}}^*$ under the extended rule that satisfies $\hat{\bm{\theta}}^*>\bm{\theta}^*$.

Under the second long-range extension with $\gamma_r=0$ and $\gamma_p>0$, a rating profile $\hat{\bm{\theta}}^*$ is an equilibrium rating profile iff:
\begin{eqnarray*}
& \!\!\!\!\!\!\!\!\!\!\!\!\!\!\!\!\!\!\!\!\!\!\!\!\!\!\!\!\!\!\!\!\!\!\!\!\!\!\!\!\!\!\!\!\!\!\!\!\!\!\!\!\!\!\!\!\!\!\!\! 0 \leq -(1-\delta_i) c_i^\prime(\hat{e}_i^*) + \delta_i \alpha_i \mu q_i^\prime(\hat{e}_i^*) \cdot \\ 
& \left[ \frac{b_i(d(\bm{\hat{\theta}}^*)_{k_i})-b_i(d(\bm{\hat{\theta}}^*)_{k_i+1}}{d(\bm{\hat{\theta}}^*)_{k_i}-d(\bm{\hat{\theta}}^*)_{k_i+1}} + \gamma_p \frac{b_i(d(\bm{\hat{\theta}}^*)_{k_i+1})-b_i(d(\bm{\hat{\theta}}^*)_{k_i+2})}{d(\bm{\hat{\theta}}^*)_{k_i}-d(\bm{\hat{\theta}}^*)_{k_i+1}} \right], 
\end{eqnarray*}
when $k_i=1$,
\begin{eqnarray*}
& \!\!\!\!\!\!\!\!\!\!\!\!\!\!\!\!\!\!\!\!\!\!\!\!\!\!\!\!\!\!\!\!\!\!\!\!\!\!\!\!\!\!\!\!\!\!\!\!\!\!\!\!\!\!\!\!\!\!\!\! 0 \leq -(1-\delta_i) c_i^\prime(\hat{e}_i^*) + \delta_i \alpha_i \mu q_i^\prime(\hat{e}_i^*) \cdot \\ 
& \left[ \frac{b_i(d(\bm{\hat{\theta}}^*)_{k_i})-b_i(d(\bm{\hat{\theta}}^*)_{k_i+1}}{d(\bm{\hat{\theta}}^*)_{k_i}-d(\bm{\hat{\theta}}^*)_{k_i+1}} + \gamma_p \frac{b_i(d(\bm{\hat{\theta}}^*)_{k_i+1})-b_i(d(\bm{\hat{\theta}}^*)_{k_i+2})}{d(\bm{\hat{\theta}}^*)_{k_i}-d(\bm{\hat{\theta}}^*)_{k_i+1}} \right], \\
& \!\!\!\!\!\! 0 \geq -(1-\delta_i) c_i^\prime(\hat{e}_i^*) + \delta_i \alpha_i \mu q_i^\prime(\hat{e}_i^*) \cdot \frac{b_i(d(\bm{\hat{\theta}}^*)_{k_i-1})-b_i(d(\bm{\hat{\theta}}^*)_{k_i}}{d(\bm{\hat{\theta}}^*)_{k_i-1}-d(\bm{\hat{\theta}}^*)_{k_i}}, 
\end{eqnarray*}
when $k_i\in\{2,\ldots,N-1\}$, and
\begin{eqnarray*}
& \!\!\!\!\!\! 0 \geq -(1-\delta_i) c_i^\prime(\hat{e}_i^*) + \delta_i \alpha_i \mu q_i^\prime(\hat{e}_i^*) \cdot \frac{b_i(d(\bm{\hat{\theta}}^*)_{k_i-1})-b_i(d(\bm{\hat{\theta}}^*)_{k_i}}{d(\bm{\hat{\theta}}^*)_{k_i-1}-d(\bm{\hat{\theta}}^*)_{k_i}},
\end{eqnarray*}
when $k_i=N$.

Following similar reasoning as in the first extension, we can find an equilibrium rating profile $\hat{\bm{\theta}}^*$ under the second extended rule that satisfies $\hat{\bm{\theta}}^*>\bm{\theta}^*$. \hfill $\Box$

\end{document}